\documentclass[preprint,prb,superscriptaddress,amssymb,amsmath,floatfix,citeautoscript]{revtex4-1}
\usepackage[pdftex]{graphicx}

\begin{document}

\title{Experimental metrology to obtain thermal phonon transmission coefficients at solid interfaces}
\author{Chengyun Hua\textsuperscript{a}}
\author{Xiangwen Chen\footnote{\text{
C. Hua and X. Chen contributed equally to this work.}}}
\author{Navaneetha K. Ravichandran}
\author{Austin J. Minnich\footnote{\text{
To whom correspondence should be addressed. E-mail: aminnich@caltech.edu}}}
\affiliation{%
 Division of Engineering and Applied Science\\
 California Institute of Technology, Pasadena, California 91125,USA
}%

\date{\today}

\begin{abstract}

Interfaces play an essential role in phonon-mediated heat conduction in solids, impacting applications ranging from thermoelectric waste heat recovery to heat dissipation in electronics. From the microscopic perspective, interfacial phonon transport is described by transmission coefficients that link vibrational modes in the materials composing the interface. However, direct experimental determination of these coefficients is challenging because most experiments provide a mode-averaged interface conductance that obscures the microscopic detail. Here, we report a metrology to extract thermal phonon transmission coefficients at solid interfaces using ab-initio phonon transport modeling and a thermal characterization technique, time-domain thermoreflectance. In combination with transmission electron microscopy characterization of the interface, our approach allows us to link  the atomic structure of an interface to the spectral content of the heat crossing it. Our work provides a useful perspective on the microscopic processes governing interfacial heat conduction. 

\end{abstract}

\pacs{}
\maketitle

\clearpage

\section{Introduction}\label{sec:introduction}

Interfaces play an essential role in phonon-mediated heat conduction in solids.\cite{swartz_thermal_1989, Cahill2014Review} Material discontinuities lead to thermal phonon reflections that are manifested on a macroscopic scale as a thermal boundary resistance (TBR), also called Kapitza resistance, $R_k$, that relates the temperature drop at the interface to the heat flux flowing across it. TBR exists at the interface between any dissimilar materials due to differences in phonon states on each side of the interface.\cite{Swartz1989} Typical interfaces often possess defects or roughness which can lead to additional phonon reflections and hence higher TBR. 

TBR plays an increasingly important role in applications, particularly as device sizes decrease below the intrinsic mean free paths (MFPs) of thermal phonons.\cite{Cahill2014Review} At sufficiently small length scales, TBR can dominate the total thermal resistance. For instance, the effective thermal conductivity of a superlattice can be orders of magnitude smaller than that of the constituent materials due to TBR.\cite{pettersson_theory_1990, Chen1998,Ravichandran2014,Chen2013} This physical effect has been used to realize thermoelectrics with high efficiency\cite{SciencePaper,Biswas2012} and dense solids with exceptionally low thermal conductivity\cite{Chiritescu2007}. On the other hand, TBR can lead to significant thermal management problems\cite{Pop2010,Moore2014,Cho2015} in applications such as LEDs\cite{Su2012,Han2013} and high power electronics\cite{Yan2011,Cho2015}.

Numerous works over several decades have investigated the microscopic origin of TBR at solid-solid interfaces, starting with studies performed at low temperatures ($\sim 1 $ K), in which heat is carried predominantly by phonons with frequencies less than 1 THz.\cite{klitsner_phonon_1987, Swartz1987} At these low temperatures and for pristine, ordered interfaces, transmission coefficients can be obtained from continuum elastic theory in an analogy with Snell's law for light; this model is known as the acoustic mismatch model (AMM).\cite{Khalatnikov1952,Little1959} The AMM was shown to explain the experimentally measured values of TBR at various solid-solid interfaces.\cite{Swartz1987} At higher temperatures (above 1 K), heat transport across the interfaces was found to be sensitive to surface roughness. For the limit of completely diffuse scattering in which transmitted and reflected phonons cannot be distinguished, Swartz constructed the diffuse mismatch model (DMM).\cite{swartz_thermal_1989} Despite the success of these models at explaining TBR at low temperatures, they generally fail at temperatures larger than 40 K and are unable to account for the atomistic structure of the interface.

Recent works have focused on remedying these deficiencies. Optical methods enable the routine measurement of TBR over a wide range of temperatures for various metal-dielectric interfaces \cite{Lyeo2006,Norris2009,Cheaito2015,Schmidt2010,duda_role_2010} as well as at interfaces with variable bonding strength. \cite{OBrien2012,Losego2012} Other works have examined the temperature dependence of the thermal conductivity\cite{Wang2011} in nanocrystalline samples. Computational atomistic methods such as molecular dynamics\cite{Maiti1997,Stevens2007,Landry2009,IhChoi2012,Jones2013,Yang2013,Merabia2014,Liang2014,Schelling2002b} and atomistic Green's functions\cite{Zhang2006, Li2012b,Tian2014,Huang2010,Hopkins2009} have been extensively applied to obtain the transmission coefficients at interfaces with realistic atomic structure. These calculations generally predict the coefficients to decrease with increasing phonon frequency due to reflections of short wavelength phonons by atomistic roughness, a trend that is supported by basic wave physics and indirectly by experiment.\cite{Wang2011,Wilson2014b} However, a direct determination of the spectral transmission coefficients at an actual interface has not yet been reported. 

Here, we report a metrology to extract the thermal phonon transmission coefficients at a solid interface. Our approach, based on combining experimental observations with ab-initio phonon transport modeling, exploits quasiballistic transport near the interface to significantly narrow the possible transmission coefficient profiles at a solid interface compared to the bounds obtained from traditional approaches. Applying our approach in conjunction with transmission electron microscopy (TEM), we are able to directly link atomic structure to the spectral content of heat crossing the interface. Our approach is a useful tool to elucidate the microscopic transport properties of thermal phonons at solid interfaces. 

\section{Overview of approach}

Our approach is based on interpreting data from the TDTR experiment with ab-initio phonon transport model. Briefly, TDTR is a widely used optical pump-probe technique that is used to characterize thermal properties. In this experiment, a sample consists of a metal transducer film on a substrate. A pulsed laser beam from an ultrafast oscillator is split into a pump and a probe beam. The pump pulse train is modulated at a frequency from 1 to 15 MHz to enable lock-in detection, and is then used to impulsively heat the metal film coated on the sample. The transient temperature decay $Z(t)$ at the surface is detected as a change in optical reflectance by the probe beam.\cite{Capinski1996} 

In the traditional TDTR approach, this transient signal is related to the desired thermal properties by a macroscopic transfer function based on a multilayer heat diffusion model.\cite{Schmidt2008,Cahill2014Review} This function maps thermal properties such as substrate thermal conductivity and metal-substrate interface conductance to the TDTR signal, and thus these properties are obtained by varying these parameters until the simulated results match the measured data sets. Put another way, one must solve an inverse problem that links the data sets to the unknown parameters; this calculation is often performed using a nonlinear least squares algorithm.

This approach is widely used and has provided important insights into a wide range of metal-semiconductor interfaces \cite{}. A drawback, however, is that the microscopic information about the interface is averaged into a single number, the interface conductance, obscuring the microscopic detail. Hopkins et al used TDTR measurements on a variety of metal films with varying phonon cutoff frequencies to extract spectral information about phonon transmission\cite{Cheaito2015}. However, determining transmission coefficients is still challenging due to variations in phonon density of states in each metal. 

In this work, we aim to directly extract the transmission coefficients from TDTR data by replacing the macroscopic transfer function based on Fourier's law with a microscopic transfer function based on ab-initio phonon transport modeling. Just as in the traditional approach, we seek to identify the parameters that best fit the TDTR data sets. Although this fitting process is much more complex than the traditional approach, in principle it is the same widely-used procedure.

\begin{figure*}[t!]
\centering
\includegraphics[scale = 0.75]{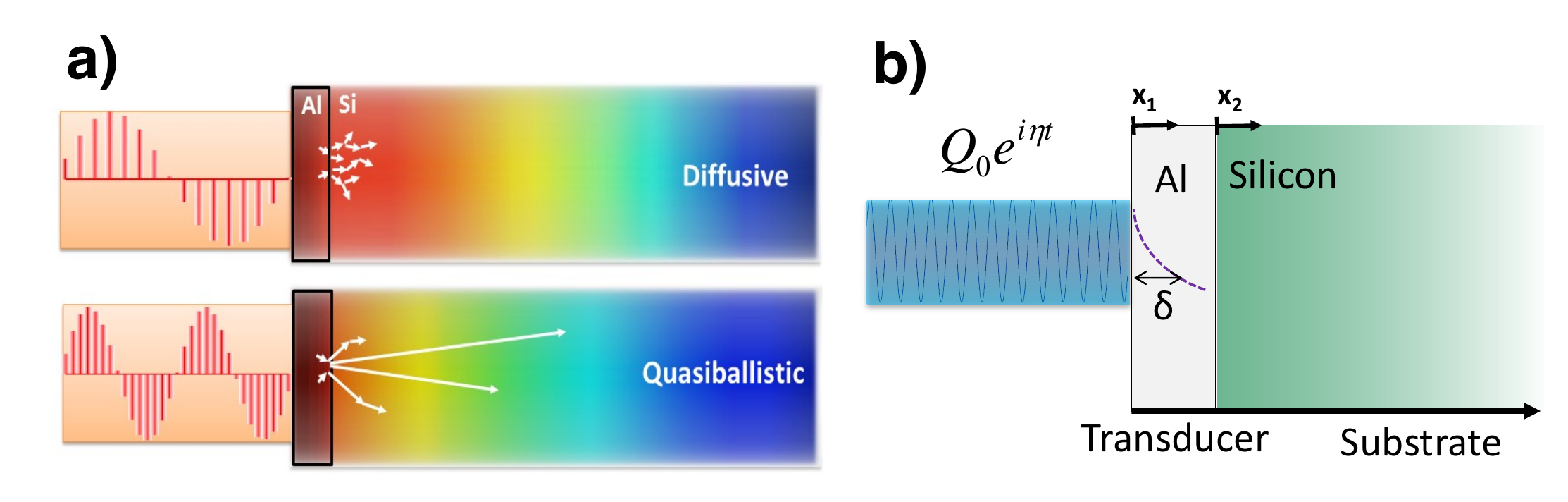}
\caption[Schematic of the principle underlying the measurement of transmission coefficients]{(a) Schematic of the principle underlying the measurement of transmission coefficients. If the characteristic length scale of the thermal transport is much longer than the phonon MFPs, information about the interfacial distribution is lost due to strong scattering. If some MFPs are comparable to the thermal length scale, the non-equilibrium distribution at the interface propagates into the substrate where it can be detected. (b) 2D schematic of the experimental configuration subject to a modulated heating source: a double layer structure of a transducer film on a substrate (sample). $Q_0$ is the amplitude of the heating source, $\eta$ is angular modulation frequency, $\delta$ is the optical penetration depth of the heating source, and $x$ is the cross-plane transport direction. $x_1$ and $x_2$ correspond to the coordinate systems used in transducer and substrate accordingly.}
\label{fig:Schematics}
\end{figure*}

There are several conditions that must be satisfied to successfully extract transmission coefficients using this approach. First, it is essential that part of the non-equilibrium phonon distribution emerging from the interface propagate into the substrate ballistically. As illustrated in Fig.~\ref{fig:Schematics} (a), when MFPs are much shorter than the characteristic length scale of the thermal gradient, information about the phonon distribution at the interface is lost due to scattering. In this case, the TDTR signal is largely insensitive to the transmission coefficient profile so long as the overall interface conductance remains fixed. 

On the other hand, if some phonons have sufficiently long MFPs, the non-equilibrium phonon distribution penetrates into the substrate. In this case, as we will show in subsequent sections, the TDTR signal depends not only on the magnitude of the interface conductance but on the spectral profile of the transmission coefficients. It is this sensitivity that we will exploit to retrieve the transmission coefficients from the TDTR data sets. 

This discussion implies that not every substrate will be suitable for our approach as sufficiently long phonons MFPs compared to the induced thermal gradient are required. Fortunately, many experimental reports have demonstrated clear evidence of this quasiballistic heat transport regime in different material systems \cite{Koh2007,Siemens2010,Minnich2011a,Regner2012,Johnson2013,Vermeersch2015a}. In particular, MFPs in Si are generally accepted to exceed one micron at room temperature. \cite{Cuffe2014} Considering that thermal penetration depths in Si are on the same order in TDTR \cite{}, Si is a suitable substrate for this work.

Second, we must determine the microscopic transfer function that maps the transmission coefficients directly to the TDTR signal without any artificial fitting parameters. This step is challenging due to the difficulty of efficiently solving the BTE for the TDTR experiment. A number of simplified models\cite{Maznve2011,Minnich2011b,Wilson2013,Vermeersch2015a,Regner2014,Koh2014,Maassen2015} have been proposed, but these models make various approximations that limit their predictive capability. In this work, we overcome this challenge using two recent advances we reported for rigorously solving the spectral BTE under the relaxation time approximation (RTA) that yield a factor of $10^4$ speedup compared to existing methods and allows the first ab-initio phonon transport modeling of TDTR free of artificial parameters or simplifications of the phonon dispersion. This solution is derived in the next section.

Before deriving the necessary functions, we briefly summarize the necessary steps in our approach. First, we obtain TDTR data for a given metal/substrate sample. We then calculate the intrinsic phonon dispersion and lifetimes for each material that are then inserted into the ab-initio transfer function derived in the next section. Finally, the transmission coefficients are varied until the optimized transmission profiles are identified that best match the experimental data. Thus, the procedure is identical to that used in traditional TDTR measurements excepting the complications of deriving and inverting the microscopic transfer function.

\section{Derivation of transfer function}\label{sec:Modeling}

We now describe the derivation of the microscopic transfer function $Z(T_{12}(\omega), t)$ that maps transmission coefficients to the TDTR amplitude and phase data as a function of time. The result of the derivation is a function for which the only inputs are the phonon dispersions and lifetimes for each material composing the interface, and the only unknown parameters are the spectral transmission coefficients $T_{12}(\omega)$. The output of the function is the TDTR amplitude and phase signal versus delay time. For this work, the phonon dispersion and lifetimes for Si were calculated from first-principles with no adjustable parameters by Lucas Lindsay. We have also used first-principles inputs from Mingo \emph{et al}. with no appreciable difference in results or conclusions. 

Let us first briefly review the signal formation in TDTR. Since the thermal response given by the BTE is a linear time-invariant system, the output transient signal $Z(t)$ of TDTR can be represented in terms of frequency response solution through the following equation\cite{Schmidt2008}
\begin{equation}\label{eq:TDTRSignal}
Z(T_{12}(\omega), t) = \sum\limits_{n = -\infty}^{\infty}H(T_{12}(\omega), \eta_0+n\eta_s)e^{in\eta_st},
\end{equation}
where $\eta_0$ is the reference angular frequency of the periodic heating, $\eta_s$ is the angular sampling frequency set by the repetition rate of the laser pulses, and $H(T_{12}(\omega), \eta)$ is surface temperature response subject to a periodic heating at frequency $\eta$ given the transmission coefficients $T_{12}(\omega)$. Therefore, identifying $Z(T_{12}(\omega), t)$ requires computing $H(T_{12}(\omega), \eta)$, or the surface temperature frequency response to a periodic heating in a double-layer structure of a metal film on a substrate as shown in Fig.~\ref{fig:Schematics} (b).

Thermal transport in an isotropic crystal, assuming only cross-plane heat conduction, is described by the one-dimensional (1D) spectral Boltzmann transport equation (BTE) under the relaxation time approximation (RTA)\cite{Majumdar1993},
\begin{eqnarray}\label{eq:BTE}
\frac{\partial g_{\omega}}{\partial t} &+& \mu v_{\omega} \frac{\partial g_{\omega}}{\partial x} = -\frac{g_{\omega}+f_0(T_0)-f_0(T)}{\tau_{\omega}}+\frac{Q_{\omega}}{4\pi} \\
f_0(T) &=& \frac{1}{4\pi}\hbar \omega D(\omega) f_{BE}(T) \approx f_0(T_0)+\frac{1}{4\pi}C_{\omega}\Delta T,
\label{eq:BEDist_Linearized}
\end{eqnarray}
where $g_{\omega} =\hbar\omega D(\omega)[f_{\omega}(x,t,\mu)-f_0(T_0)]$ is the deviational distribution function, $f_0 = f_0(x,t)$ is the equilibrium distribution function, $\mu = cos(\theta)$ is the directional cosine, $v_{\omega}$ is the phonon group velocity, $\tau_{\omega}$ is the phonon relaxation time, and $Q_{\omega}(x,t)$ is the spectral volumetric heat generation.  Assuming a small temperature rise, $\Delta T = T - T_0$, relative to a reference temperature, $T_0$, the equilibrium distribution is proportional to $\Delta T$, as shown in Eq.~(\ref{eq:BEDist_Linearized}).  Here, $\hbar$ is the reduced Planck constant, $\omega$ is the phonon frequency, $D(\omega)$ is the phonon density of states, $f_{BE}$ is the Bose-Einstein distribution, and $C_{\omega} = \hbar\omega D(\omega)\frac{\partial f_{BE}}{\partial T}$ is the mode specific heat. The volumetric heat capacity is then given by $C = \int_0^{\omega_m}C_{\omega}d\omega$ and the Fourier thermal conductivity $k = \int_0^{\omega_m}k_{\omega}d\omega$, where $k_{\omega} = \frac{1}{3} C_{\omega}v_{\omega} \Lambda_{\omega}$ and $\Lambda_{\omega} = \tau_{\omega}v_{\omega}$ is the phonon MFP. To close the problem, energy conservation is used to relate $g_{\omega}$ to $\Delta T$, given by 
\begin{equation}
\int\int_0^{\omega_m} \left[\frac{g_{\omega}(x,t)}{\tau_{\omega}}-\frac{1}{4\pi}\frac{C_{\omega}}{\tau_{\omega}}\Delta T(x,t) \right]d\omega d\Omega = 0,
\label{eq:EnergyConservation}
\end{equation}
where $\Omega$ is the solid angle in spherical coordinates and $\omega_m$ is the cut-off frequency. Note that summation over phonon branches is implied without an explicit summation sign whenever an integration over phonon frequency or MFP is performed. 

We now divide our discussion into three parts: transducer film, substrate, and interface. The BTE in the transducer film can be reformulated as a Fredholm integral equation of the second kind\cite{hua_semi-analytical_2015}; the solution in the substrate can be treated as the Green's function to the BTE.\cite{Hua2014b} The solutions in the two layers depend on each other through the interface conditions that enforce conservation of heat flux.

\subsection{Transducer film}\label{subsec:film}

The metal thin film serves as an optical transducer that absorbs the incident optical energy while also enabling the observation of temperature decay through the thermoreflectance coefficient. In our work, we neglect electrons and consider that heat is only carried by phonons in Al. Justification for this approximation is given in Appendix \ref{sec:electrons}. 

Since the system is modulated at a given frequency $\eta$, we can assume that both $g_{1\omega}$ and $\Delta T_1$ are of the form $e^{i\eta t}$ to define $g_{1\omega} = \widetilde{g}_{1\omega}(x_1,\mu)e^{i\eta t}$ and $\Delta T_1 = \Delta \widetilde{T}_1(x_1)e^{i\eta t}$. The volumetric heat generation in thin film is given by $Q_{\omega} = Q^0_{\omega}e^{i\eta t}e^{-x_1/\delta}$, where the amplitude of heating source $Q_0 = \int_0^{\omega_m} Q^0_{\omega} d\omega$. We also assume that phonons are specularly reflected at $x_1 = 0$, \emph{i.e.} $\widetilde{g}_{1\omega}(x_1=0,\mu) =\widetilde{g}_{1\omega}(x_1=0,-\mu)$. Substituting the definition of $\widetilde{g}_{1\omega}$ and $\widetilde{T}_1(x_1)$ and the specular boundary condition at $x_1 = 0$ into Eq.~(\ref{eq:BTE}) leads to a first-order ODE for $\widetilde{g}_{1\omega}(x_1,\mu)$. Its solution is given by
\begin{eqnarray}\label{eq:ODEsolution_gPlus}\nonumber
\widetilde{g}^+_{1\omega}(x_1,\mu) &=& B_{\omega}e^{-\frac{\gamma_{1\omega}}{\mu}(d+x_1)}+\int_0^{d}\frac{C_{1\omega}\Delta \widetilde{T}(x'_1)+Q^0_{\omega}e^{-x/\delta}\tau_{1\omega}}{4\pi\Lambda_{1\omega}\mu}e^{\frac{\gamma_{1\omega}}{\mu}(x'_1+x_1)}dx'_1 \\ 
&+& \int_0^{x_1}\frac{C_{1\omega}\Delta \widetilde{T}(x'_1)+Q^0_{\omega}e^{-x/\delta}\tau_{1\omega}}{4\pi\Lambda_{1\omega}\mu}e^{\frac{\gamma_{1\omega}}{\mu}(x'_1-x_1)}dx'_1\ (\mu \in (0, 1]) \\ \label{eq:ODEsolution_gMinus}\nonumber
\widetilde{g}^-_{1\omega}(x_1,\mu) &=& B_{\omega}e^{\frac{\gamma_{1\omega}}{\mu}(d-x_1)} \\
&-& \int_x^d\frac{C_{1\omega}\Delta \widetilde{T}(x'_1) + Q^0_{\omega}e^{-x/\delta}\tau_{1\omega}}{4\pi\Lambda_{1\omega}\mu}e^{\frac{\gamma_{1\omega}}{\mu}(x'_1-x_1)}dx'_1\ (\mu \in [-1, 0]),
\end{eqnarray}
where $\gamma_{1\omega} = (1+i\eta\tau_{1\omega})/\Lambda_{1\omega}$, $d$ is the film thickness, and $B_{\omega}$ are the unknown coefficients determined by the interface condition at $x_1 = d$. Here, $\widetilde{g}_{1\omega}^+(x_1,\mu)$ indicates the forward-going phonons and $\widetilde{g}_{1\omega}^-(x_1,\mu)$ the backward-going phonons. 

To close the problem, we plug Eqs.~(\ref{eq:ODEsolution_gPlus}) \& (\ref{eq:ODEsolution_gMinus}) into Eq.~(\ref{eq:EnergyConservation}) and obtain an integral equation for temperature as:
\begin{equation}\label{eq:FilmTemperature}
\Delta \widetilde{T}(\widehat{x}_1) - \int_0^1\Delta \widetilde{T}(\widehat{x}'_1)K(\widehat{x}'_1,\widehat{x}_1)d\widehat{x}'_1 = \int_0^{\omega_m}B_{\omega}F^1_{\omega}(\widehat{x}_1)d\omega + F^2(\widehat{x}_1),
\end{equation}
where $\widehat{x}_1 = x_1/d$. The kernel function $K(\widehat{x}'_1,\widehat{x})$ is given by
\begin{equation}\label{eq:Kernel}
K(\widehat{x}'_1,\widehat{x}_1) = \frac{1}{2\int_0^{\omega_m}\frac{C_{1\omega}}{\tau_{1\omega}}d\omega}\int_0^{\omega_m}\frac{C_{1\omega}}{\tau_{1\omega}\text{Kn}_{1\omega}}\{E_1[\widehat{\gamma}_{1\omega}(\widehat{x}'_1+\widehat{x}_1)]+E_1[\widehat{\gamma}_{1\omega}|\widehat{x}'_1-\widehat{x}_1|]\}d\omega
\end{equation}
and the two inhomogeneous functions are given by
\begin{eqnarray}\label{eq:InhomoF1}
F^1_{\omega}(\widehat{x}_1) &=& \frac{2\pi}{\int_0^{\omega_m}\frac{C_{1\omega}}{\tau_{1\omega}}d\omega}\frac{1}{\tau_{1\omega}}\{E_2[\widehat{\gamma}_{1\omega}(1+\widehat{x}_1)]+E_2[\widehat{\gamma}_{1\omega}(1-\widehat{x}_1)]\}\\ \label{eq:InhomoF2}\nonumber
F^2(\widehat{x}_1) &=&  \frac{2\pi}{\int_0^{\omega_m}\frac{C_{1\omega}}{\tau_{1\omega}}d\omega}\int_0^1\int_0^{\omega_m}\frac{Q^0_{\omega}e^{-\rho\widehat{x}'_1}}{\text{Kn}_{1\omega}}\{E_1[\widehat{\gamma}_{1\omega}(\widehat{x}'_1+\widehat{x}_1)]+E_1[\widehat{\gamma}_{1\omega}|\widehat{x}'_1-\widehat{x}_1|]\}d\omega d\widehat{x}'\\
\end{eqnarray}
where Kn$_{1\omega} = \Lambda_{1\omega}/d$ is the Knudsen number, $\widehat{\gamma}_{1\omega} =\frac{1+i\eta\tau_{1\omega}}{\text{Kn}_{1\omega}}$, and $E_n(x)$ is the exponential integral given by\cite{Gangbook}:
\begin{equation}\label{eq:ExpInt}
E_n(x) = \int_0^1 \mu^{n-2}e^{-\frac{x}{\mu}}d\mu.
\end{equation}

Recently, we have developed a spectral method to efficiently solve Eq.~(\ref{eq:FilmTemperature}) in Ref. 60\nocite{hua_semi-analytical_2015}. Briefly, the functions in Eq.~(\ref{eq:FilmTemperature}) can be expanded as a finite cosine series, such as
\begin{equation}\label{eq:TempExp}
\Delta \widetilde{T}_{1(N)}(\widehat{x}_1) \approx \sum\limits_{n = 0}^{N}t_n \text{cos}(n\pi\widehat{x}_1)
\end{equation}
and
\begin{eqnarray}\label{eq:KernelExp}\nonumber
K_{(N)}(\widehat{x},\widehat{x}')=\frac{1}{4}k_{00}&+&\frac{1}{2}\sum\limits_{m=1}^{N}k_{m0}cos(m\pi \widehat{x})+\frac{1}{2}\sum\limits_{n=1}^{N}k_{0n}cos(n\pi \widehat{x}'), \\
&+&\sum\limits_{m=1}^N\sum\limits_{n=1}^N k_{mn}cos(m\pi \widehat{x})cos(n\pi \widehat{x}')
\end{eqnarray}
where $N$ is the truncated basis number, and $t_n$'s and $k_{nm}$'s are the Fourier coefficient. Similarly, $F^1_{\omega}(\widehat{x}_1)$ and $F^2(\widehat{x}_1)$ are also expanded in term of cosines. Following the steps in the above reference, we can express the temperature as
\begin{equation}\label{eq:FilmTemp}
\Delta \widetilde{T}_{1}(\widehat{x}_1) = [\underline{\underline{A}}^{-1}(\underline{\underline{f_1}}\underline{B}+\underline{f_2})]^{T}\underline{\phi (x)}
\end{equation}
where the matrix $\underline{\underline{A}}$ contains elements $A_{00}=1-\frac{k_{00}}{4}$, $A_{0n}=-\frac{1}{2}k_{0n}$, $A_{n0}=-\frac{k_{n0}}{4}$, $A_{nn}=1-\frac{1}{2}k_{nn}$ and $A_{nm}=-\frac{1}{2}k_{nm}$ ($m\neq n \neq 0$) and $\underline{B}$ is a $N_{\omega}$ column vector of the unknown coefficients $B_{\omega}$, where $N_{\omega}$ is the number of discretization in phonon frequency. $\underline{\underline{f_1}}$ is a $N \times N_{\omega}$ matrix, consisting of the Fourier coefficients of $F^1_{\omega_i}(\widehat{x}_1)$ evaluated at each phonon frequency $\omega_i$ and $\underline{f_2}$ is a N column vector, consisting of the Fourier coefficients of $F^2(\widehat{x}_1)$.

Then, $\widetilde{g}^+_{1\omega}(x_1,\mu)$ and $\widetilde{g}^-_{1\omega}(x_1,\mu)$ can be expressed in terms of the unknown coefficients $B_{\omega}$ by plugging Eq.~(\ref{eq:FilmTemp}) into Eqs.~(\ref{eq:ODEsolution_gPlus}) and (\ref{eq:ODEsolution_gMinus}).

\subsection{Substrate}\label{subsec:substrate}

The substrate can be treated as a semi-infinite region subject to a surface heat flux. Therefore, the BTE for the substrate becomes
\begin{equation}\label{eq:SubstrateBTE}
i\eta\widetilde{g}_{2\omega}+ \mu v_{2\omega} \frac{\partial \widetilde{g}_{2\omega}}{\partial x_2} = -\frac{\widetilde{g}_{2\omega}}{\tau_{2\omega}}+\frac{C_{2\omega}}{4\pi \tau_{2\omega}}\Delta \widetilde{T}(x_2) + \frac{1}{2}P_{\omega}v_{2\omega}|\mu | \delta (x_2),
\end{equation}
where the unknown coefficients $P_{\omega}$'s are determined through the interface conditions. 

We then apply the Green's function method given in Ref. 61\nocite{Hua2014b}. The unknown distribution function in spatial frequency domain is then written as 
\begin{equation}\label{eq:SubstrateDistribution}
\widetilde{g}_{2\omega}(\eta,\xi_2)=\frac{C_{2\omega}}{4\pi}\frac{\Delta \widetilde{T}_2(\eta,\xi_2)+\frac{1}{2}P_{\omega}\Lambda_{2\omega}|\mu|/C_{2\omega}}{1+i\eta\tau_{2\omega}+i\mu\xi_2\Lambda_{2\omega}},
\end{equation}
and the temperature profile 
\begin{equation}\label{eq:SubstrateTemp}
\Delta \widetilde{T}_2(\eta,\xi_2) = \frac{\int_0^{\omega_m}P_{\omega}v_{2\omega}\frac{1+i\eta \tau_{2\omega}}{(\Lambda_{2\omega}\xi_2)^2}\log\left[1+\left(\frac{\Lambda_{2\omega}\xi_2}{1+i\eta\tau_{2\omega}}\right)^2\right]d\omega}{\int_0^{\omega_m}\frac{C_{2\omega}}{2\pi\tau_{2\omega}}\left[1-\frac{1}{\Lambda_{2\omega}\xi_2}\text{tan}^{-1}\left(\frac{\Lambda_{\omega}\xi_2}{1+i\eta\tau_{2\omega}}\right)\right]d\omega},
\end{equation} 
where $\xi_2$ is the Fourier variable of $x_2$.

Again, to express $\widetilde{g}_{2\omega}$ only in terms of unknown coefficients $P_{\omega}$, we simply plug Eq.~(\ref{eq:SubstrateTemp}) into Eq.~(\ref{eq:SubstrateDistribution}).

\subsection{Interface condition}\label{sec:InterfaceCondition}

The unknown coefficients in the solutions of transducer film and substrate are obtained by applying appropriate interface conditions. Here, we use the elastic transmission interface condition with mode conversion, closely following the work by Minnich \emph{et al}.\cite{Minnich2011b} Briefly, for a given mode $i$, the heat fluxes outgoing from the interface, $q_{1\omega}^{i-}$ and $q_{2\omega}^{i+}$, must be equal to the reflected and transmitted heat fluxes incident to the interface, $q_{1\omega}^{i+}$ and $q_{2\omega}^{i-}$. By assuming elastic and diffuse scattering, the transmission and reflection process for each phonon frequency is treated independently and the heat flux equality condition must be satisfied for each frequency and polarization. 

The interface conditions are
\begin{eqnarray}\label{eq:IC_1}\nonumber
\int_0^1 g_{2\omega}^{i+} v^i_{2\omega}\mu d\mu &=& \sum_{j}T^{ji}_{12}(\omega)\int_0^1 g_{1\omega}^{j+} v^j_{1\omega}\mu d\mu + \sum_{j}R^{ji}_{21}(\omega) \int_0^1 g_{2\omega}^{j-}v^j_{2\omega}\mu d\mu, \\  \\ \nonumber
\label{eq:IC_2}
\int_0^1 g_{1\omega}^{i-} v^i_{1\omega}\mu d\mu &=& \sum_{j}T^{ji}_{21}(\omega)\int_0^1 g_{2\omega}^{j-} v^j_{2\omega}\mu d\mu+\sum_{j}R^{ji}_{12}(\omega)\int_0^1 g_{1\omega}^{j+} v^j_{1\omega}\mu d\mu, \\
\end{eqnarray}
where $T^{ji}_{12}(\omega)$ is the transmission coefficient of mode $j$ at frequency $\omega$ from side 1 to side 2 as mode $i$, $R^{ji}_{21}(\omega)$ is the reflection coefficient of  mode $j$ at frequency $\omega$ from side 2 back into side 2 as mode $i$, and so on.

The next question is how $T^{ij}_{12}(\omega)$ is related to the other reflection and transmission coefficients. The reflection coefficients are related to the transmission coefficients by energy conservation given by 
\begin{equation}
\sum_j T^{ij}_{12}(\omega)+R^{ij}_{12}(\omega) = 1,
\end{equation}
and
\begin{equation}
\sum_j T^{ij}_{21}(\omega)+R^{ij}_{21}(\omega) = 1.
\end{equation}
$T^{ji}_{21}(\omega)$ is related to $T^{ij}_{12}(\omega)$ through the principle of detailed balance, which requires that no net heat flux can transmit across the interface when both materials are at an equilibrium temperature $T$. Applying this condition to every phonon mode on each side of the interface for each polarization and frequency gives:
\begin{equation}\label{eq:DetailedBalance}
T^{ij}_{12} (\omega)C^{i}_{1\omega}v^{i}_{1\omega} = T^{ji}_{21} (\omega) C^{j}_{2\omega}v^{j}_{2\omega}.
\end{equation}
Therefore, we need to specify $T^{ij}_{12}(\omega)$, $R^{ij}_{12}(\omega)$, and $R^{ij}_{21}(\omega)$. 

Let us first consider a special case where no mode conversion is allowed ($T^{ij}_{12}(\omega)$, $T^{ij}_{21}(\omega)$, $R^{ij}_{12}(\omega)$, $R^{ij}_{21}(\omega) = 0$ for $i \neq j$). Then, the interface conditions become
\begin{eqnarray}\label{eq:IC_1_special}
\int_0^1 g_{2\omega}^{i+} v_{2\omega}\mu d\mu &=& T^{ii}_{12}(\omega)\int_0^1 g_{1\omega}^{i+} v^i_{1\omega}\mu d\mu + R^{ii}_{21}(\omega) \int_0^1 g_{2\omega}^{i-}v^{i}_{2\omega}\mu d\mu, \\
\label{eq:IC_2_special}
\int_0^1 g_{1\omega}^{i-} v_{1\omega}\mu d\mu &=& T^{ii}_{21}(\omega)\int_0^1 g_{2\omega}^{i-} v^i_{2\omega}\mu d\mu+R^{ii}_{12}(\omega)\int_0^1 g_{1\omega}^{i+} v^{i}_{1\omega}\mu d\mu,
\end{eqnarray}
and the detail balance becomes
\begin{equation}\label{eq:DetailedBalance_special}
T^{ii}_{12} (\omega)C^{i}_{1\omega}v^{i}_{1\omega} = T^{ii}_{21} (\omega) C^{i}_{2\omega}v^{i}_{2\omega}.
\end{equation}
Therefore, once $T^{ii}_{12}(\omega)$ is specified, all the other transmission and reflection coefficients are determined. For now, we only consider this special case and neglect the mode conversion in our BTE simulations. Later, we show that the mode specific transmission coefficients cannot be resolved by the TDTR measurements and the measurable quantity is $\sum_{j}T^{ij}_{12}(\omega)$ instead of individual transmission coefficients.  For simplicity, we will use $T_{12}(\omega,p) $ rather than the summation.  

\subsection{Justification of approximations}\label{sec:ApproxJustification}

In this section, we justify the approximations made in the derivation of the transfer function. First, we have neglected the role of electrons, either in carrying heat in the metal or in carrying heat across the interface by direct coupling to substrate phonons. For the first point, in Appendix \ref{sec:electrons}, we solve the BTE explicitly including electron-phonon coupling and show that it has negligible effect on the TDTR signal and hence the transmission coefficients. For the second point, our approach cannot rule out this mechanism. However, the available evidence in the literature \cite{Lyeo2006, lee_phonon_2013,giri_mechanisms_2015} shows that the metal electron-substrate phonon coupling effect is too small to be observed for several material systems. Therefore, little evidence exists to support a large contribution from this mechanism. As a result, we assume that heat is carried across the interface solely by phonons.

The second assumption is the neglect of mode conversion, or the change of phonon polarization as phonons transmit or reflect at the interface. In Appendix \ref{sec:ModeConversion}, we explicitly calculate the transfer function including this mechanism, showing that it has very little impact on the results. Therefore, fitting the data with or without mode conversion yields identical results.

\subsection{Overview of calculation}

We now describe how these pieces fit together to provide the transfer function $Z(T_{12}(\omega),t)$. At any given frequency $\eta$, the analytical expression of the unknown distributions from both sides, $g_{1\omega}^{\pm}$ and $g_{2\omega}^{\pm}$ can be obtained according to Sec.~\ref{subsec:film} and Sec.~\ref{subsec:substrate}, which then are evaluated at $x_1 = d$ and at $x_2 = 0$, respectively. Given the values of $T_{12}(\omega)$, the values of $R_{12}(\omega)$, $T_{21}(\omega)$ and $R_{21}(\omega)$ can be inferred, and the unknown coefficients $P_{\omega}$ and $B_{\omega}$'s are obtained by plugging Eqs.~(\ref{eq:ODEsolution_gPlus}), (\ref{eq:ODEsolution_gMinus}), and (\ref{eq:SubstrateDistribution}) into Eqs.~(\ref{eq:IC_1_special}) and (\ref{eq:IC_2_special}) and solving the linear system.

Once $B_{\omega}$ is known, the temperature profile in the transducer film is solved using Eq.~(\ref{eq:FilmTemp}) as well as the surface temperature $H(\eta)$. At a given modulation frequency $\omega_0$, $\eta$ is chosen to be $\omega_0+n\omega_L$, where integer $n$ is typically ranging from -50 to 50. Using Eq.~(\ref{eq:TDTRSignal}) yields $Z(T_{12}(\omega), t)$, which is directly compared to experimental data $Z_{exp}(t)$.

\section{Sensitivity of TDTR signal to transmission coefficients}

With the transfer function obtained, it is useful to re-examine the conditions that are required to successfully extract transmission coefficients from TDTR data.  First, the number of unknowns should be on the order of the number of data points. As an estimate, consider that we take TDTR data at five modulation frequencies. As in Eq.~(\ref{eq:TDTRSignal}), in the frequency domain each TDTR signal consists of surface temperature responses at different frequencies, the lowest of which is the modulation frequency. Typically, among those surface temperature responses, the responses at the lowest four to five frequencies contain most of the thermal information.\cite{Cahill2004} With around five TDTR data sets containing amplitude and phases at five different modulation frequencies, we obtain 40-50 unique data points.

Now consider the number of unknowns. If we take the materials to be isotropic, the transmission coefficients depend only on phonon frequency. In our typical discretization, we find that there around one hundred transmission coefficients that must be fit. However, these coefficients are not all independent due to a smoothness constraint - physically, the transmission profile cannot fluctuate arbitrarily, an intuition supported by atomistic simulations \cite{Li:2012g,Li:2012j}. Qualitatively, this requirement implies that each coefficient depends on the adjacent coefficients, decreasing the effective number of unknowns by a value on the order of two to three. Therefore, the number of unknowns is comparable to the number of data points.

\begin{figure*}[t!]
\centering
\includegraphics[scale = 0.32]{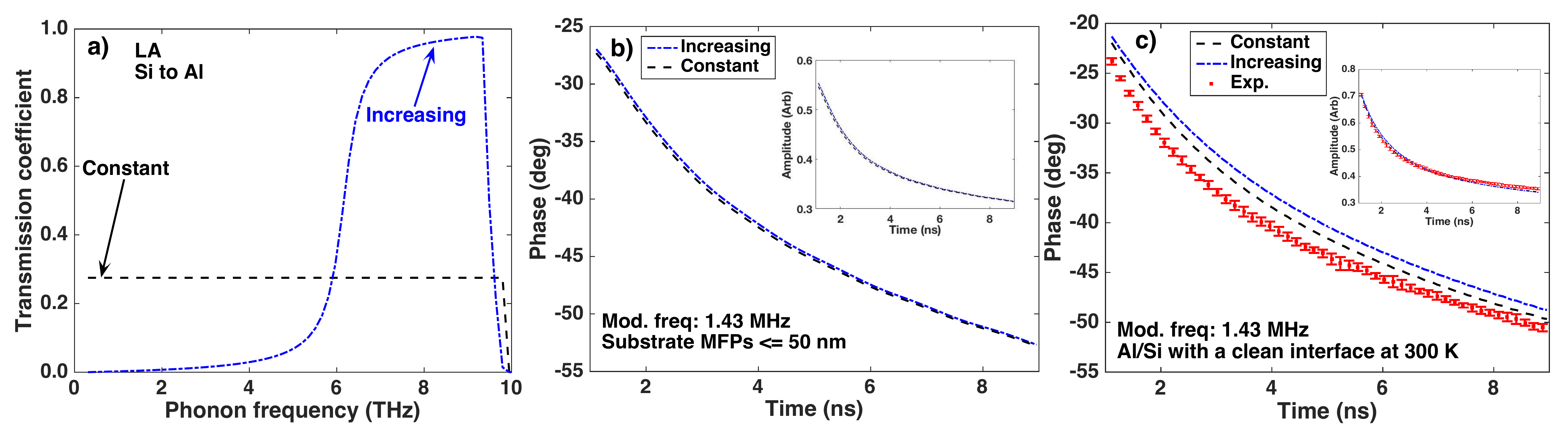}
\caption[Sensitivity of TDTR signals]{(a) Two artificial transmission coefficient profiles from Si to Al for the longitudinal branch. The two profiles have different trend: a constant value (dashed line) and an increasing trend (dotted-dash line), but both give the same interface conductance. (b) The amplitude of the simulated TDTR signals using a constant transmission coefficient (dashed line) and a increasing transmission coefficient profile (dotted-dash line) with Silicon's dispersion and a cut-off scattering rates such that there no MFP exceeding 50 nm. (c) The amplitude of the measured TDTR signal (solid line) for Al/Si at 300 K and the simulated TDTR signals using the transmission coefficient profiles shown in (a). Insets: the phase of the corresponding TDTR signals. The two signals are almost identical under diffusion transport, while, in the quasiballistic regime, the two simulated TDTR signals are no longer identical and do not match with the measured signal either.}
\label{fig:TDTR_sensitivity}
\end{figure*}

Ample data points are a necessary but not sufficient condition to enable extraction of transmission coefficients. The last requirement is that the TDTR signal should be sensitive to the shape of the transmission coefficient profile. In the case of heat diffusion, this requirement is not satisfied: so long as the interface conductance is unchanged, the TDTR signal will not change because scattering in the substrate obscures the interfacial phonon distribution. To demonstrate this point, we simulated TDTR signals using two different transmission coefficient profiles as shown in Fig.~\ref{fig:TDTR_sensitivity} (a) with the scattering rates of Si modified such that no MFP exceeds 50 nm. The chosen transmission profiles possess the same interface conductance. The amplitude and phase of the simulated TDTR signals for these two transmission coefficient profiles are shown in Fig.~\ref{fig:TDTR_sensitivity} (b). The figure shows that they are nearly identical even though the two transmission coefficient profiles are completely different, demonstrating that in the diffusion regime the TDTR signal only depends on the magnitude of interface conductance. 

On the other hand, in the quasiballistic regime the TDTR signal is sensitive to the shape of the transmission coefficient profile, enabling the coefficients to be obtained with tight constraints. We demonstrate this sensitivity in Fig.~\ref{fig:TDTR_sensitivity} (c). For this calculation, the same transmission coefficient profiles are used but the bulk scattering rates of Si from Lindsay et al are used without modification. The longest MFP in silicon is on the order of micrometers, \cite{Broido2007,Cuffe2014} and the characteristic length scale of a TDTR experiment with a modulation frequency around 1 MHz is also on the order of a micron. Therefore, the transport in the TDTR experiment for Al/Si is quasiballistic. In this case, the TDTR signals using two different transmission coefficient profiles are no longer identical as shown in Fig.~\ref{fig:TDTR_sensitivity} (c), demonstrating that in the quasiballistic regime the TDTR signal is sensitive to both the magnitude of interface conductance and the spectral profile. 

Fig.~\ref{fig:TDTR_sensitivity} (c) plots a measured TDTR signal. From the plot, one can immediately tell the transmission coefficient profiles used in the calculations are not correct as the experimental and calculation signals do not match. The correct transmission coefficient profile will reproduce the measured TDTR signal. Our procedure to identify this profile is described in the next section.

\section{Solution of inverse problem}\label{subsec:TransmissionMeasurements}

The final step is to solve the inverse problem that identifies the transmission coefficients that best explain the observed data. From the BTE model, we obtain a surface temperature decay curve as a function of time just like the one measured in the experiments. For a given sample, the actual transmission coefficient profile as a function of phonon frequency will minimize the difference between the simulation curves and experimental TDTR traces at all modulation frequencies. 

We solve the inverse problem using a particle swarm optimization (PSO) method to search for the optimal profile. The goal of the PSO method is to minimize the objective function defined as
\begin{equation}
f = \alpha|g_{ab-initio}(T_{12}(\omega)) - g_{measured}| + (1-\alpha)\int\left(\frac{d^2T_{12}}{d\omega^2}\right)^2d\omega. \label{eq:ObjectiveFunction}
\end{equation}
The first part of the equation evaluates the norm of the difference between the experimentally measured and BTE-simulated TDTR signals given a transmission profile profile $T_{12}(\omega)$. The second part of the equation evaluates the second derivative of the transmission coefficient profile, serving as the smoothness penalty function. Note that the smoothness of the profiles is the only constraint we impose in the objective function. The smoothing parameter $\alpha$ determines the relative importance of the second part to the first part.  If $\alpha = 1$, then no smoothness constraint is imposed. Here, we use
\begin{equation}
\alpha = \frac{\int\left(\frac{d^2T^{0}_{12}}{d\omega^2}\right)^2d\omega}{|g_{ab-initio}(T^{0}_{12}(\omega)) - g_{measured}|},
\end{equation}
where $T^0_{12}(\omega)$ is the initial profile. The formula is chosen such that the first and second parts of the equation have the same order of magnitude. 

To search for the optimal profile that minimizes the objective function, the PSO algorithm randomly initializes a collection of transmission coefficient profiles and evolves them in steps throughout the phase space which contains all possible transmission coefficient profiles. At each step and for each profile, the algorithm evaluates the objective function defined as above. After this evaluation, the algorithm decides how each profile should evolve according to the current best profile. The profile evolves, then the algorithm reevaluates. The algorithm stops when the objective function reaches the desired value. The transmission coefficient profile that achieves the minimum value of the objective function is the optimal profile that explains the data.

However, since the inverse problem is ill-posed, a unique solution does not exist. We generate a probability density plot for the transmission coefficients using Gibbs sampling to explore adjacent regions of the optimal transmission coefficient profile. We first randomly generate around 1000 profiles by perturbing the optimal profile with a smooth function defined using the following formula
\begin{equation}
\delta = A[r_1\text{cos}(2\pi\omega/\omega_{max}r_2+2\pi r_3)+r_4\text{sin}(2\pi\omega/\omega_{max}r_5+2\pi r_6)], \label{eq:perturbation}
\end{equation} 
where the amplitude of the perturbation $A$ is 0.1, and $r_1$, $r_2$, $r_3$, $r_4$, $r_5$ and $r_6$ are random numbers between 0 to 1. We evaluate the objective function at all the perturbed profiles and recorded the values. Then, we start the Gibbs sampling process. At each iteration, we randomly draw a profile, $a$, from the stored population and compare the value of its corresponding objective function, $f_{n}$ to the one from the previous step, $f_{n-1}$ evaluated at profile $b$. If $f_{n}$ is less than $f_{n-1}$, we accept $a$ and kept $f_{n}$. If not, a random number $r$ is drawn and compared to $u = p/(1+p)$, where 
\begin{equation}
p = \text{exp}\left(\frac{f_{n}-f_{n-1}}{T_0}\right). \label{eq:GibbsSampling}
\end{equation}
If $r$ is smaller than $u$, then we accept $a$ and kept $f_{n}$. If not, we reject $a$ and update $f_{n}$ to be $f_{n-1}$.  The system temperature, $T_0$, is chosen such that the stationary distribution is gradually changing. Here, $T_0$ is set to be the mean value of the objective functions of all the perturbed samples. We keep track of how many times each profile was chosen at each iteration and generated a histogram of the occurrence frequency of each profile. We stop the sampling process when the histogram becomes stationary. This occurrence frequency is also called the likelihood of the transmission coefficient profiles. The higher the value of a profile's likelihood is, the better the fit with the experimentally measured TDTR signals at different modulation frequencies. Thus by combining the PSO method with the Gibbs sampling algorithm, we are able to determine the most likely transmission coefficients at the interface between Si and Al.

\section{Results}

\subsection{Phonon transmission coefficients}
\begin{figure*}[t!]
\centering
\includegraphics[scale = 0.55]{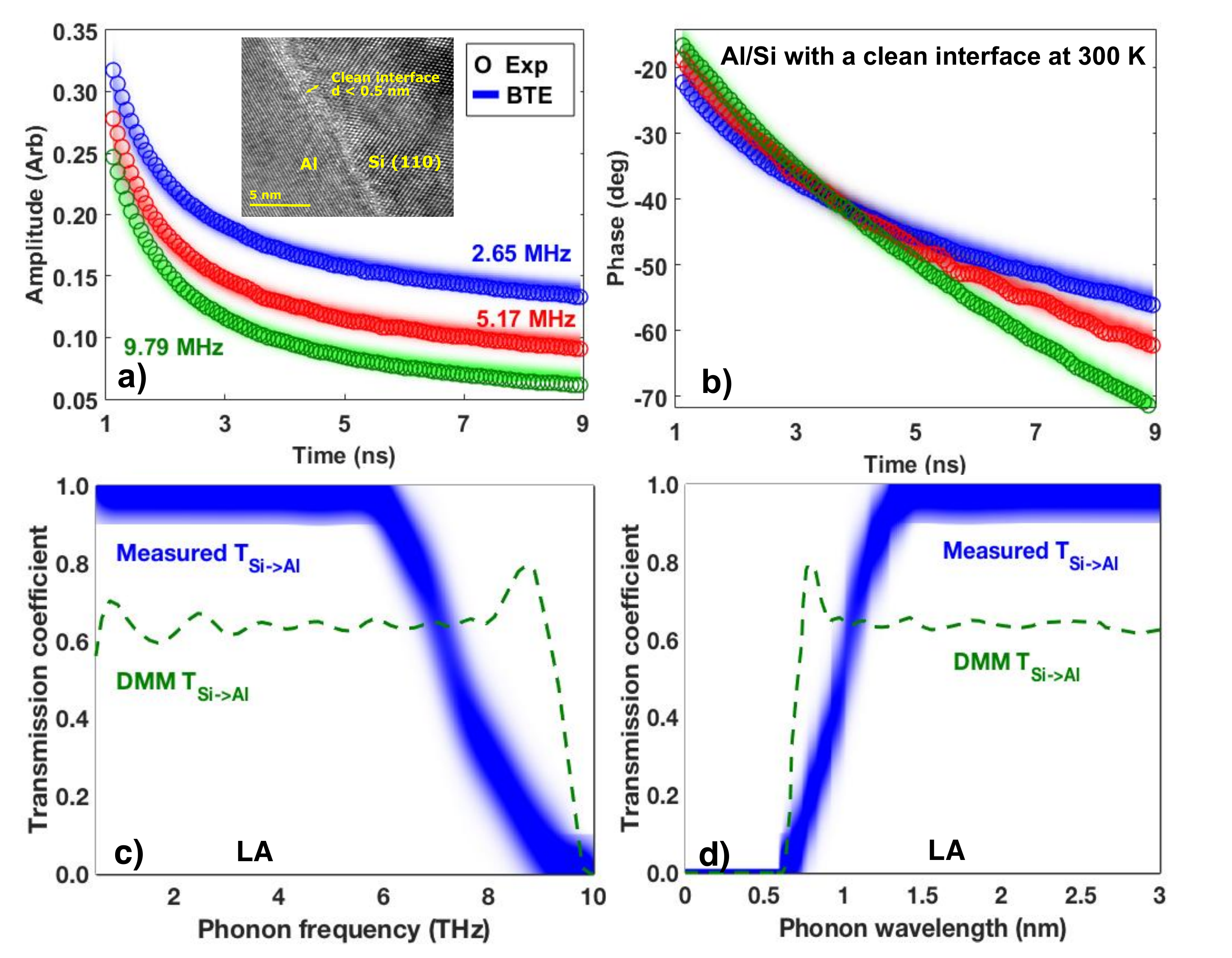}
\caption[Measurements and simulations of TDTR experiments on Al/Si with a clean interface]{Experimental TDTR data (symbols) on this sample at $T = 300$ K for modulation frequencies $f = 2.68$, $5.51$ and $9.79$ MHz along with the (a) amplitude and (b) phase fit to the data from the BTE simulations (shaded regions), demonstrating excellent agreement between simulation and experiment. The shaded stripes denoted BTE simulations correspond to the likelihood of the measured transmission coefficients possessing a certain value as plotted. Inset: TEM image showing the clean interface of an Al/Si sample with the native oxide removed. The interface thickness is less than $0.5$ nm. Transmission coefficients of longitudinal phonons T$_{\text{Si} \rightarrow \text{Al}}(\omega)$ (blue shaded region) versus (c) phonon frequency and (d) phonon wavelength, along with the DMM transmission coefficient profile (green dashed line) that gives the same interface conductance as the measured T$_{\text{Si} \rightarrow \text{Al}}(\omega)$. The intensity of the shaded region corresponds to the likelihood that the transmission coefficient possesses a given value.}
\label{fig:TDTRSignal_clean_300K}
\end{figure*}

We demonstrate our transmission coefficient measurements on an Al film on Si substrate with the native oxide removed by Hydrofluoric acid prior to Al deposition, yielding a clean interface. The TEM image in Fig.~\ref{fig:TDTRSignal_clean_300K} (a) shows the interface thickness is less than 0.5 nm. The amplitude and phase of signals from the lock-in amplifier at different modulation frequencies are given in Fig.~\ref{fig:TDTRSignal_clean_300K}. For reference, solving the usual inverse problem with the macroscopic transfer function on this data set yields $G \approx 280$ MW/m$^2$-K and $k \approx 140$ W/m-K, in good agreement with prior works and literature values for the thermal conductivity of Si.\cite{Minnich2011a,Wilson2014b} Although the good agreement is often taken as evidence that the macroscopic transfer function is valid for Si, this conclusion is incompatible with several independent ab-initio calculations that clearly show that heat is carried by phonons with MFPs exceeding the thermal penetration depth of TDTR.\cite{Broido2007,Esfarjani2011} This prediction has recently been experimentally confirmed by Cuffe \emph{et al} using thermal measurements on variable thickness silicon membranes.\cite{Cuffe2014} This fact implies that quasiballistic transport should be readily observable in a typical TDTR experiment on Si, despite the seemingly correct thermal properties measured. This apparent contradiction is resolved by observing that the signal measured in TDTR strongly depends on the spectral profile of the transmission coefficients in the quasiballistic regime as shown in Fig.~ \ref{fig:TDTR_sensitivity}.

We represent the transmission coefficient as a probability density plot, with the color intensity indicating the likelihood that a single transmission coefficient curve passing through a particular point at a given phonon frequency is able to simultaneously explain all of the data in Fig.~\ref{fig:TDTRSignal_clean_300K}, without any other adjustable parameters. The result is shown in Fig.~\ref{fig:TDTRSignal_clean_300K}(c). The figure shows that the transmission coefficient from Si to Al for longitudinal phonons, T$_{\text{Si} \rightarrow \text{Al}}(\omega)$, starts at unity, its maximum possible value, and decreases steadily to near zero for high phonon frequencies ($\sim 10$ THz). The transmission coefficient profiles for the other polarizations have similar shapes, and so throughout the paper we plot only the longitudinal transmission coefficients for simplicity. The transmission coefficients from Al to Si, T$_{\text{Al} \rightarrow \text{Si}}(\omega)$ are calculated by satisfying the principle of detailed balance; the relationship between T$_{\text{Si} \rightarrow \text{Al}}(\omega)$ and T$_{\text{Al} \rightarrow \text{Si}}(\omega)$ reflects the differences in density of states and group velocity between the two materials. The transmission coefficients for each side of the interface and for the other polarizations are given in Appendix~\ref{sec:TransmissionCoeff}.   

\subsection{Comparison with conventional models}\label{sec:WrongModels}

\begin{figure*}
\centering
\includegraphics[scale = 0.45]{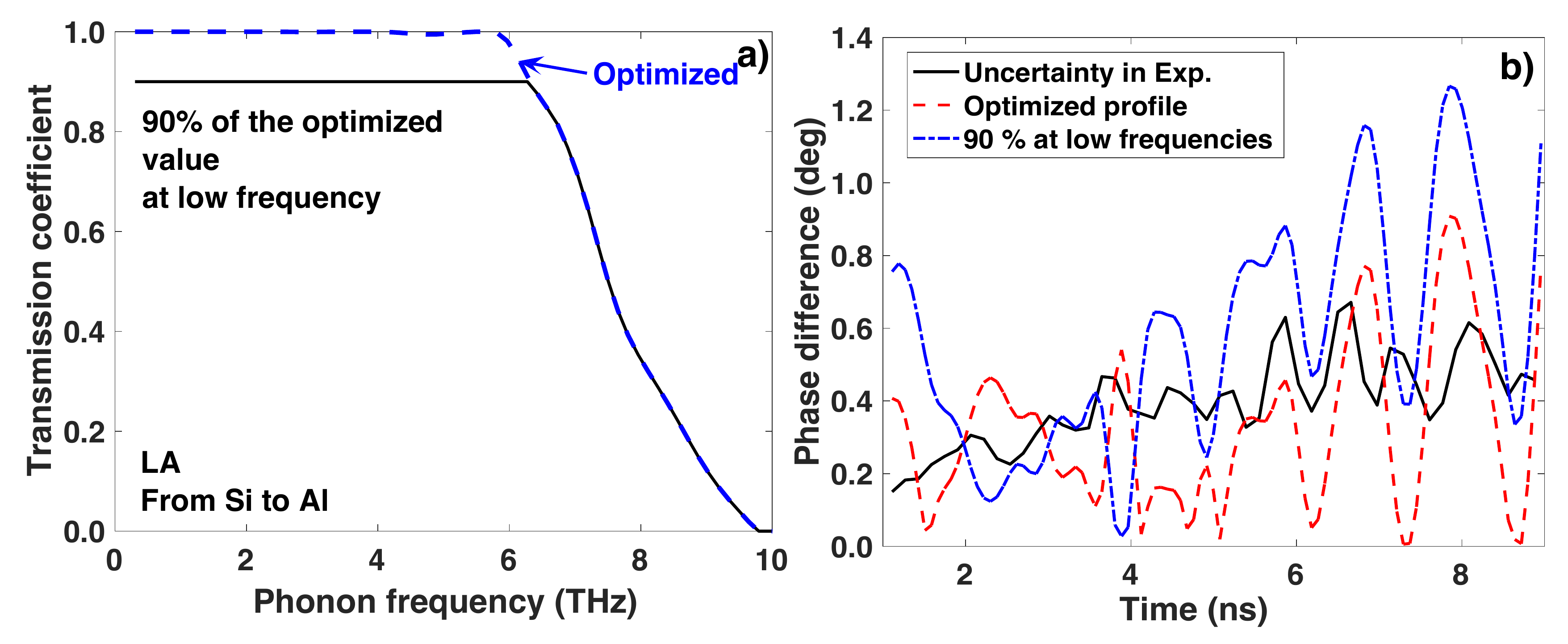}
\caption[Justification for high transmission coeffficients at low frequency modes]{(a) Optimized transmission coefficient profile (dashed line) and the profile with 90 \% reduced values at phonon frequencies less than 6 THz (solid line) versus phonon frequency for longitudinal modes at a clean Al/Si interface. (b) The absoluate phase difference between the TDTR signal and experimental data as a function of time using the optimized transmission coefficient profile (dashed line) and the profile with 90 \% reduced values at low phonon frequencies (dotted-dash line) compared to the experimental uncertainties (solid line). }
\label{fig:LowFreqJustification}
\end{figure*}

Our measured transmission coefficient profile thus indicates that longitudinal phonons with frequencies less than 6 THz are transmitted to the maximum extent allowed by the principle of detailed balance, while longitudinal phonons with frequencies larger than 8 THz are nearly completely reflected at the interface. We now examine this result in context with the common models for transmission coefficients. The AMM is often cited as an appropriate model for transmission coefficients at sufficiently low phonon frequencies. Calculating the AMM transmission coefficients for normal incidence, we obtain a value of 0.95, which is quite consistent with our observation.  Fig.~\ref{fig:LowFreqJustification} shows how the TDTR signal varies as the low phonon frequency transmission coefficients are reduced. Below 90 \% transmission, the computed signal is outside of the error bounds. Therefore, the low frequency transmission must exceed this value to match the data. 

For short wavelength phonons, the DMM would be expected to apply. At the highest phonon frequencies (shortest wavelengths), the DMM correctly predicts the trend of the measured transmission coefficients tending to zero. However, for most of the phonon spectrum, the DMM is inconsistent with our measurement.

We provide additional evidence for the inadequacy of conventional models to explain our measurements by considering two models: the gray model in which the transmission coefficient is a constant, independent of phonon frequency, and the diffuse mismatch model (DMM). The DMM is only determined by the phonon properties of the materials, such as density of states and phonon group velocity. The constant transmissivity value is chosen to yield an interface conductance $G = 284$ MW/m$^2$-K using the formula of Ref. 54\nocite{Minnich2011b}. The measured value for the clean interface is $280 \pm 10 $ MW/m$^2$-K.

\begin{figure*}[htp]
\centering
\includegraphics[scale = 0.5]{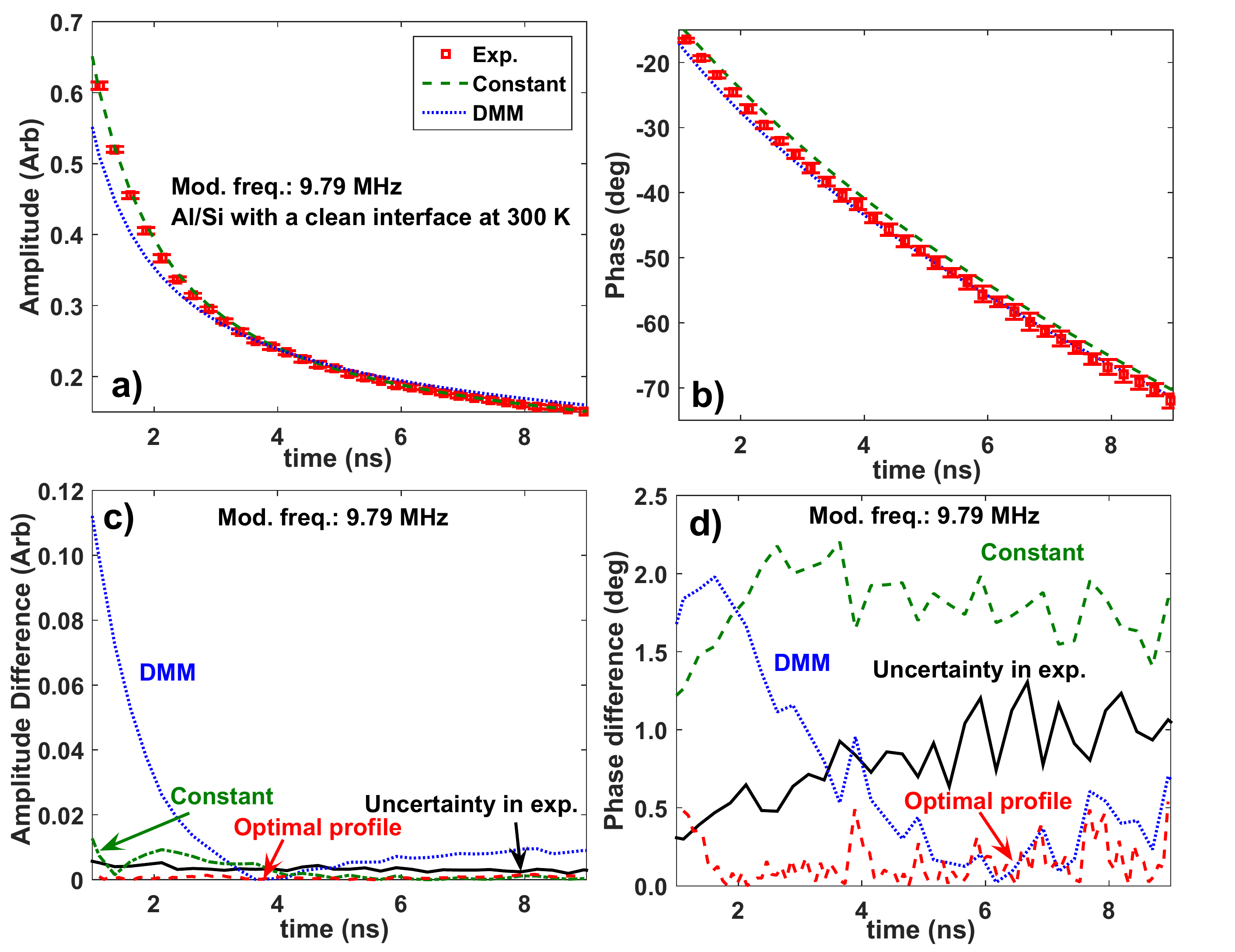}
\caption[Experimental TDTR data on Al/Si with a clean interface at 300 K for modulation frequency $f = 9.79$ MHz compared to the data from the BTE simulations using constant T$_{Si \rightarrow Al}$ and DMM]{Experimental TDTR data (symbols) on Al/Si with a clean interface at 300 K for modulation frequency $f = 9.79$ MHz along with the (a) amplitude and (b) phase compared to the data from the BTE simulations using constant T$_{Si \rightarrow Al}$ (dash-dotted lines) and DMM (dotted lines). (c) Amplitude and (d) phase difference between averaged experimental data and the BTE simulations using constant T$_{Si \rightarrow Al}$ (dash-dotted lines), DMM (dotted lines), and the optimal profile in Fig. 2 of the main text (dashed lines). The solid line indicates the uncertainty in experiments.}
\label{fig:Comparison9MHz}
\end{figure*}

Here, we demonstrate that neither of the models can explain the experimental TDTR data. As shown in Fig.~\ref{fig:Comparison9MHz}, the use of a constant transmission coefficient in the BTE model overpredicts the phase values. Similarly, the DMM underpredicts both the amplitude and phase at the early time of the signals. In Figs.~\ref{fig:Comparison9MHz} (c) \& (d), we show the deviation in amplitude and phase between the averaged experimental data at a given modulation frequency and the BTE simulations using a constant profile and DMM, demonstrating that the deviation is far beyond the uncertainty in experimental data. The uncertainty is computed by calculating the standard deviation of both amplitude and phase data for multiple runs and multiple locations on a sample. 

A better comparison for our measurements is with atomistic calculations that are not subject to the highly restrictive assumptions of the AMM and DMM. Performing this comparison, we observe that our measurements agree with numerous molecular dynamics and atomistic Green's function calculations, essentially all of which predict the general trend of decreasing transmission coefficients with increasing phonon frequency.\cite{Li2012b,Tian2014,Huang2010,Hopkins2009} In particular, our measurement of high transmission for longitudinal phonons with frequencies less than approximately 4 THz is consistent with atomistic calculations on acoustically-matched materials.\cite{Tian2014, zhao_phonon_2009} Our result also agrees with the experimental studies of polycrystalline silicon by Wang \emph{et al},\cite{Wang2011} which suggested that the transmission coefficients should decrease with increasing frequency.

\subsection{Interfacial heat flux}

\begin{figure*}[t!]
\centering
\includegraphics[scale = 0.42]{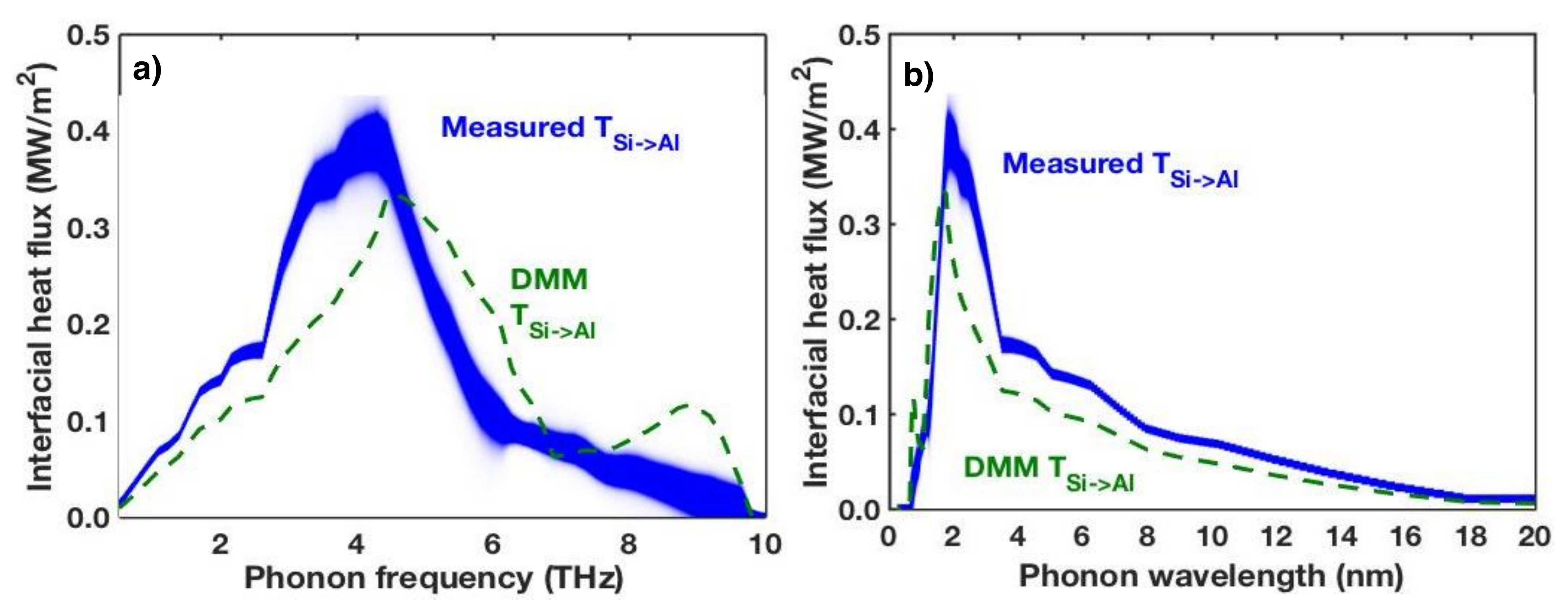}
\caption[Transmission coefficients and spectral heat flux at the interface]{Spectral heat flux with the measured (blue shaded region) and DMM (green dashed line) transmission coefficient profiles across the interface versus (a) phonon frequency and (b) phonon wavelength. Phonons with frequencies less than approximately 4 THz carry a significant amount of heat across the interface. The intensity of the shaded region reflects the likelihood of the corresponding transmission coefficients.}
\label{fig:InterfaceHeatFlux}
\end{figure*} 

Using this transmission coefficient profile, we plot the spectral interfacial heat flux versus phonon frequency and accumulative heat flux versus phonon wavelength in Fig.~\ref{fig:InterfaceHeatFlux}. Our results show that most of interfacial heat flux is carried by phonons with frequencies less than 4 THz, with the contribution from higher frequencies strongly reduced due to their small transmission coefficients. In fact, we find that the contribution of phonons with frequencies less than 4 THz is essential to explain our observations: we are unable to explain the measured data without the contribution of phonons with frequencies less than 4 THz. Similarly, we find that we can only explain the measurements using the exact phonon dispersion for Al computed from DFT; simple dispersion relations such as Debye model cannot explain the data because they underestimate the contribution of low frequency phonons to thermal transport. 

\subsection{Robustness of the measured transmission coefficients}

\begin{figure*}[t!]
\centering
\includegraphics[scale = 0.55]{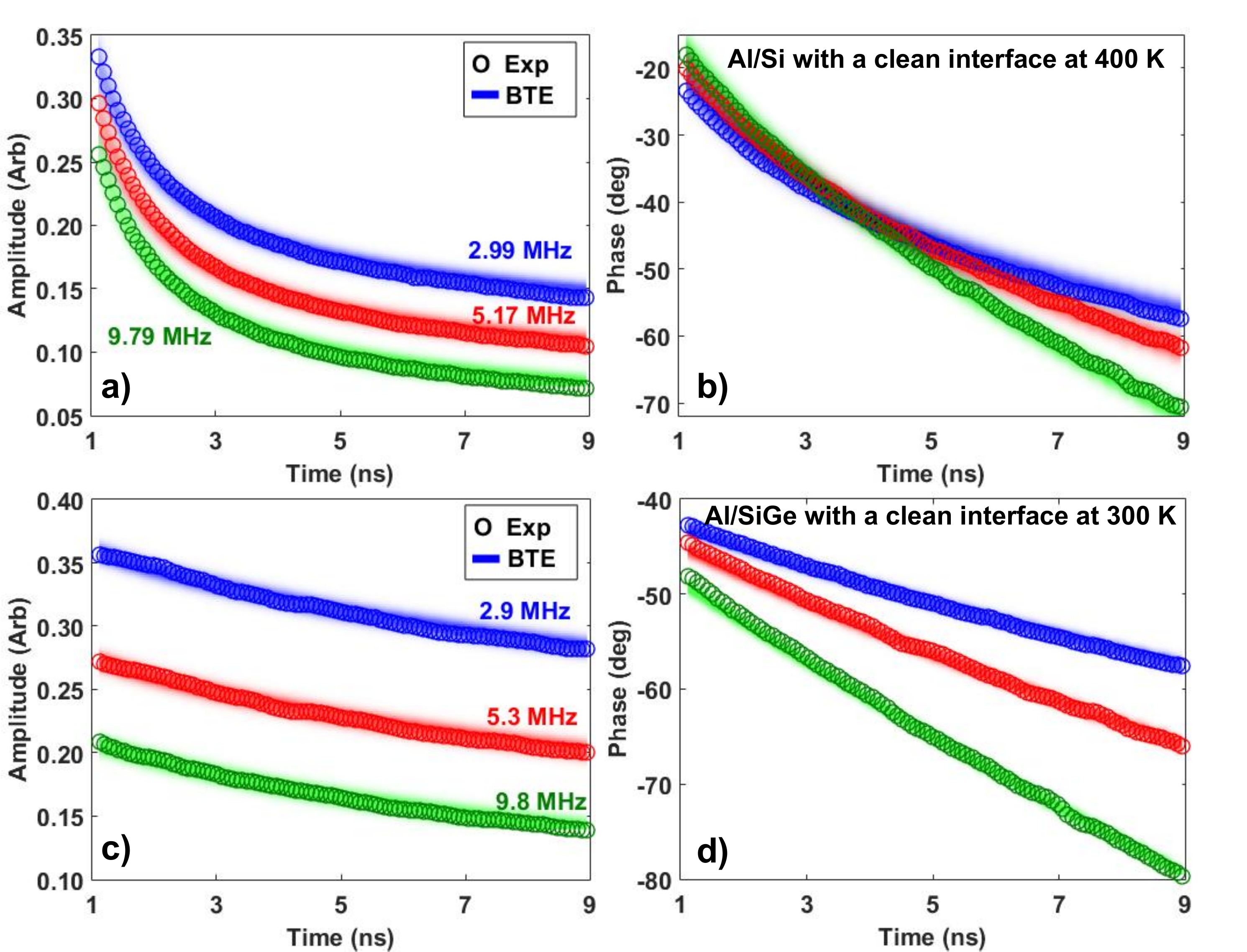}
\caption[TDTR measurements on Al/Si  at different temperatures and Al/SiGe.]{ (a) Amplitude and (b) phase as a function of time at modulation frequencies $f = 2.99$, $5.17$ and $9.79$ MHz from experiments (symbols) and simulations (shaded regions) for Al on Si with a clean interface at 400 K. (c) Amplitude and (d) phase as a function of time at modulation frequencies $f = 2.9$, $5.3$ and $9.8$ MHz from experiments (symbols) and simulations (shaded regions) for Al on SiGe with a clean interface at 300 K. The magnitude and trend of the experimental data are reproduced using the same transmission coefficient profile as in Fig.~\ref{fig:TDTRSignal_clean_300K}(c) without any adjustable parameters.}
\label{fig:TDTR_clean_400K}
\end{figure*}

We conducted several additional experiments to confirm the robustness of the measured transmission coefficients. First, since the energy transmission at the interfaces is considered elastic, the transmission coefficients in theory should be independent of temperature. We performed TDTR measurements on the same Al/Si sample at several temperatures higher than 300 K and compared the experimental results with the calculations using the same transmission coefficient profile measured at 300 K. As shown in Figs.~\ref{fig:TDTR_clean_400K}, the calculation is in excellent agreement with experimental data at 400 K using exactly the same transmission coefficient profile obtained at 300 K. Note that this comparison does not require any adjustable parameters. Additional measurements at various temperatures are given in the Supplementary Information, and all give excellent agreement. We were unable to conduct measurements at lower temperatures due to the onset of radial heat conduction that is not accounted for in our model.

Second, we measured the transmission coefficients for Al on SiGe. While this material has an additional point defect scattering mechanism compared to pure Si, we expect the transmission coefficients to be nearly the same given that the host lattice is unchanged. The details about point defect scattering in SiGe are given in Appendix \ref{appsec:SiGe}. Figs.~\ref{fig:TDTR_clean_400K} (c) \& (d) plots the amplitude and phase of the surface temperature decays at different modulation frequencies, demonstrating that the same transmission coefficient profile shown in Fig.~\ref{fig:TDTRSignal_clean_300K}(c) yields a signal that agrees well with this independent data set, again without any adjustable parameters. This result confirms that the measured transmission coefficients for Si and SiGe substrates are indeed the same. 

Interpreting the TDTR data on the SiGe with the conventional model results in a number of physical inconsistencies that are eliminated with our approach. First, the thermal conductivity of SiGe obtained with the conventional Fourier model is around 35 W/mK and varies with modulation frequency as reported previously \cite{Koh2007}. However, the actual value is around 50 W/mK (see Appendix \ref{appsec:SiGe}). Thus TDTR does not necessarily provide the actual thermal conductivity of a material.

Second, interpreting these data with the traditional Fourier model results in interface conductances for Si and SiGe differing by more than 30 \% even though the host lattice is the same \cite{Hohensee2015}. This inconsistency is removed when interpreting the data with our microscopic model as the same transmission coefficients explain both data sets. 

\section{Effect of interface atomic structure}

\begin{figure*}[ht!]
\centering
\includegraphics[scale = 0.65]{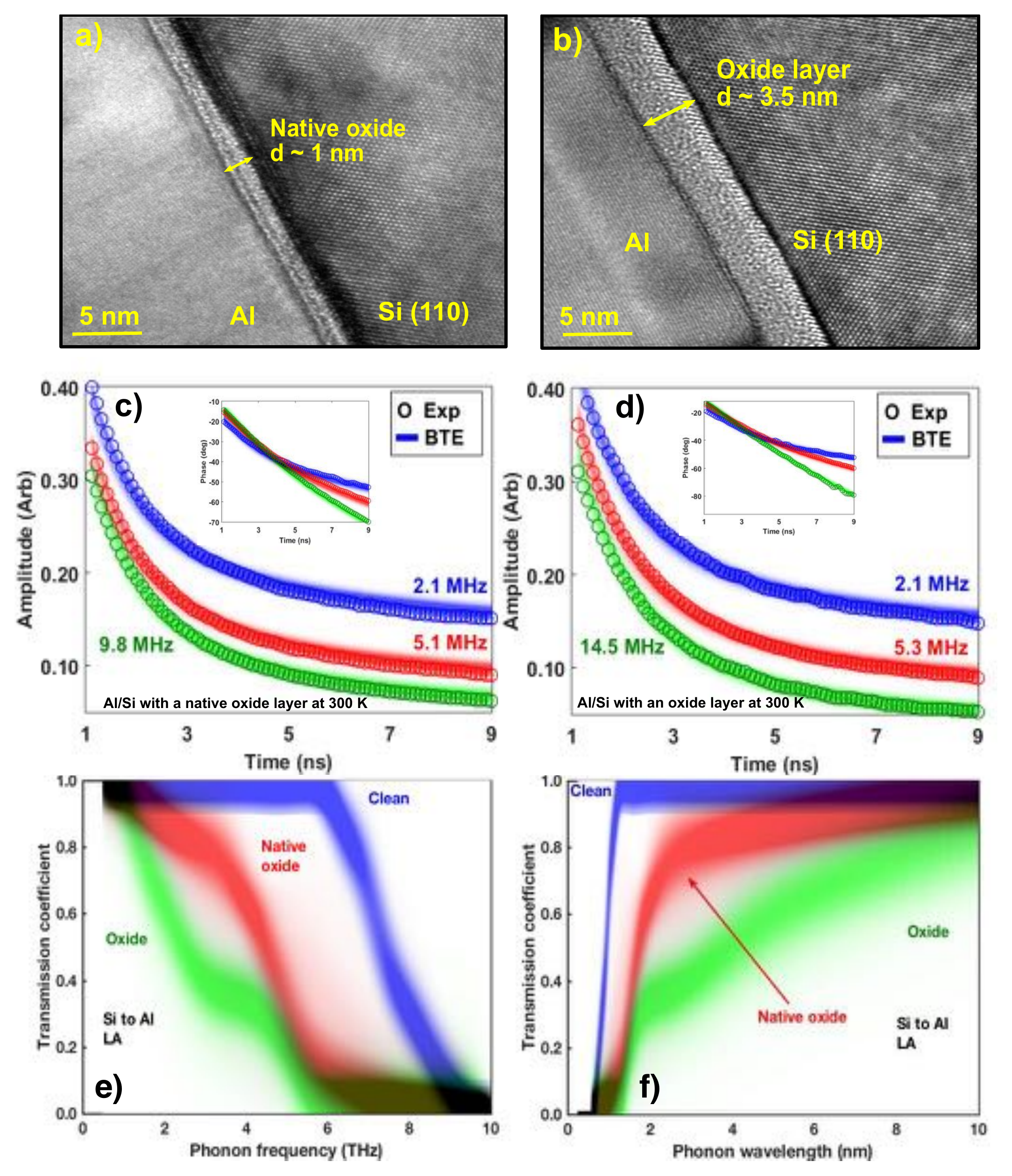}
\caption[Relationship between atomic structure and transmission coefficients.]{TEM images showing the Al/Si sample with (a) native oxide layer (thickness $\sim 1$ nm) and (b) thermally grown oxide layer (thickness $\sim 3.5$ nm).  (c) Amplitude of the surface temperature decay curves at modulation frequencies $f = 2.1$, $5.10$ and $9.80$ MHz of experiments (symbols) and simulations (shaded regions) for Al on (c) Si with native oxide layer. (d) Amplitude of the surface temperature decay curves at modulation frequencies $f = 2.1$, $5.3$ and $14.5$ MHz of experiments (symbols) and simulations (shaded regions) for Al on Si with thermally grown oxide layer. Insets: corresponding phase of the surface temperature decay curves. The corresponding transmission coefficient profiles versus (e) phonon frequency and (f) phonon wavelength show that as the interface gets rougher, phonons with frequencies less than 4 THz are more likely to be reflected.}
\label{fig:TEM_Oxide_NativeOxide}
\end{figure*}

Finally, we examine how the atomic structure of the interface affects the spectral content of the phonons carrying heat across the interface. We conducted additional measurements for Al on Si with a native oxide layer (thickness $\sim$ 1 nm as shown in a TEM image in Fig.~\ref{fig:TEM_Oxide_NativeOxide} (a)) and Si with thermally grown oxide layer (thickness $\sim$ 3.5 nm as shown in a TEM image in Fig.~\ref{fig:TEM_Oxide_NativeOxide} (b)). Since the oxide layers are sufficiently thin to neglect their thermal capacitance, we can treat them as part of the interface\cite{shen_ballistic_2014} that modifies the net transmission coefficient profile that describes transmission between Al and Si.

By solving the inverse problem with the measurement as in Figs.~\ref{fig:TEM_Oxide_NativeOxide} (c) \& (d) as input, we are able to find the transmission coefficient profiles for these two cases as shown in Figs.~\ref{fig:TEM_Oxide_NativeOxide} (e) \& (f). Compared to a clean interface, the transmission coefficients for Al on Si with a native oxide are reduced for most of the phonon modes, except those with long wavelength longer than 1 nm. When the roughness of the interface increases with a thicker oxide layer, the transmission coefficient keeps decreasing and more phonons, especially those with wavelengths between 1 and 3 nm, are reflected at the interface.  Therefore, our measurements show that phonons with wavelength shorter than the interface roughness are more likely to be reflected by the interface than phonons with wavelength longer than the interface roughness, and as the interface gets rougher, a larger fraction of the phonon spectrum is affected by the interface. In contrast to prior approaches that measure only interface conductance, here we are able to provide microscopic insight into which phonons are more likely to be reflected due to atomic-scale changes in the interface structure.

\section{Discussion}

Our work has implications for thermal metrology and technological applications. First, we have shown that TDTR is capable of providing considerable microscopic detail about thermal phonons if the measurements can be properly interpreted using a microscopic transfer function with ab-initio input. Unlike with the macroscopic transfer function, our approach provides quantitative details on the spectral content of the heat carried by phonons in the sample. As a corollary, using the macroscopic transfer function to interpret TDTR data on certain samples can lead to erroneous results. For example, the apparently correct measurement of silicon thermal conductivity is a coincidence that occurs due to the transmission coefficient profile of a typical Al/Si interface. As recently showed by Wilson et al, if the interface is modified by the introduction of an oxide layer, the measured thermal conductivity no longer coincides with the literature value. \cite{Wilson2014b} Similarly, the traditional fitting approach yields a thermal conductivity for SiGe that does not agree with its actual value measured with a bulk method. These inconsistencies are eliminated if the data are interpreted with our approach. Therefore, the conventional TDTR interpretation does not necessarily provide the actual physical properties of materials. 

Second, our measurements show that the spectral profile of transmission coefficients is essential to understanding thermal transport across interfaces. Due to a lack of knowledge about interfaces, the phonon transmission coefficients are often predicted with a variety of simple models. However, this work shows that none of these models are capable of explaining the experimental measurements. Therefore, including an accurate spectral transmission coefficient profile is essential to properly describing thermal phonon transport across interfaces. 

Third, our work provides evidence that elastic transmission of phonons across an interface is the dominant energy transmission mechanism for materials with similar phonon frequencies. Our microscopic transfer function does not incorporate electrons or inelastic scattering yet is able to explain all of the measurements we performed. This observation shows that the consideration of inelastic transmission and coupling between electrons in metals and phonons in semiconductors is not necessary to explain the TDTR data sets.  

Fourth, our results provide quantitative information into which phonons transmit across interfaces of a given atomic structure. In particular, the strong frequency dependence of the transmission coefficients can be exploited to create thermal phonon filters to selectively remove parts of phonon spectrum, analogous to optical long-pass filters. Phonons with wavelength much longer than the characteristic roughness of an interface are more likely transmitted through the interface while short-wavelength phonons are mostly reflected. Our approach provides a means to determine which of these phonon modes are transmitted or reflected and thus identify which phonon modes are filtered by the interface. 

Finally, our work demonstrates the insights into heat conduction at the atomic scale that can be obtained through the interwoven application of experimental measurements in the quasiballistic heat conduction regime, ab-initio phonon transport modeling, and electron microscopy. Through our approach, we are able to provide useful microscopic detail on the spectral content of heat cross an interface and its atomic structure. Such a capability will permit the rational understanding and control of interfacial heat transport at the atomic level, a capability that could impact numerous applications.

\section*{Acknowledgements}

The authors thank L. Lindsay, J. Carrete and N. Mingo for providing the first-principles calculations for silicon, Prof. Nathan Lewis group for the access to the ellipsometer, and the Kavli Nanoscience Institute (KNI) at Caltech for the availability of critical cleanroom facilities. X. C. thanks Melissa A. Melendes, Matthew H. Sullivan and Carol M. Garland from the KNI for fabrication assistance, and Victoria W. Dix from the Lewis group at Caltech for the help with the ellipsometer measurements. This work was sponsored in part by the National Science Foundation under Grant no. CBET 1254213, and by Boeing under the Boeing-Caltech Strategic Research \& Development Relationship Agreement.

\clearpage

\appendix
\section{Role of electrons}\label{sec:electrons}

Our simulations do not include electron-phonon coupling across the Al-Si interface or electron heat conduction in the metal film. The first approximation was justified in the text in Sec.~\ref{sec:ApproxJustification}. We justify the second approximation by performing the simulations based on an electron-Fourier/phonon-BTE model that accounts for electron conduction in the metal. Briefly, this model uses spectral phonon BTE described in Sec.~\ref{sec:Modeling} coupled with a heat diffusion equation for electrons in the Al thin films after absorption of an optical laser impulse. The coupled equations are given as following:  
\begin{eqnarray}
C_{el}\frac{\partial T_{el}}{\partial t} &=& \frac{\partial}{\partial x}\left( k_{el}\frac{\partial T_{el}}{\partial x}\right)-g(T_{el}-T_{ph}) \label{eq:electrons}\\ \nonumber
\frac{\partial g_{\omega}}{\partial t} + \mu v_{\omega} \frac{\partial_{\omega}}{\partial x} &=& -\frac{g_{\omega}+f_0(T_0)-f_0(T_{ph})}{\tau_{\omega}}+g(T_{el}-T_{ph})+\frac{Q_{\omega}(x,t)}{4\pi},\\
\label{eq:phonons}
\end{eqnarray}
where $T_{el}$ and $T_{ph}$ are the temperatures of the electrons and phonons, respectively, and $C_{el}$ and $k_{el}$ are the volumetric heat capacity and the thermal conductivity of the electrons in Al, respectively. The phonon temperature is linearly coupled to the electron temperature through the electron-phonon coupling coefficient g. The values of all the constants in Eq.~\ref{eq:electrons} are tabulated in Table~\ref{tab:constants}. This system of the equations is solved by a standard finite difference method in a two layered geometry. 

\begin{figure*}[t!]
\centering
\includegraphics[scale = 0.5]{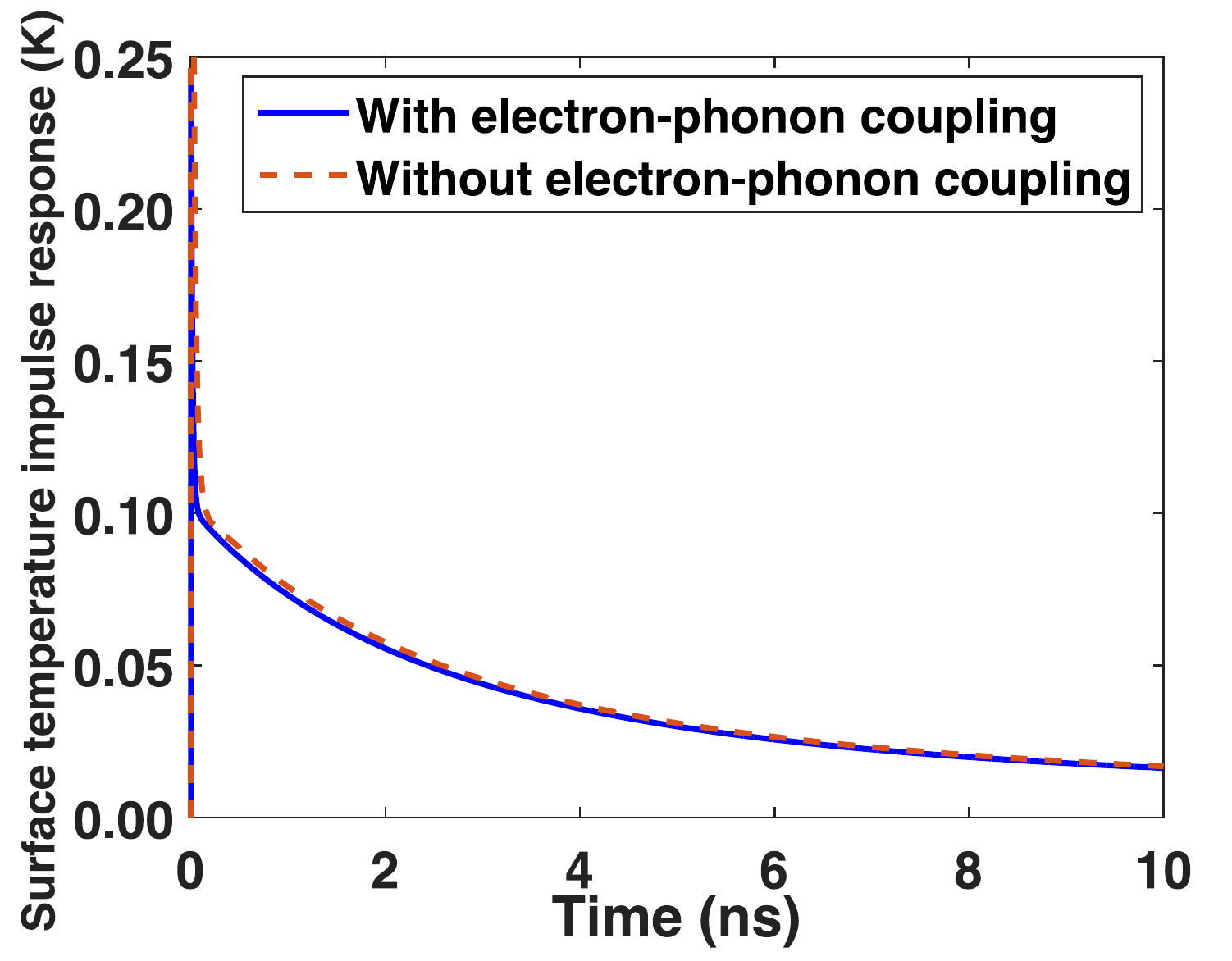}
\caption[The surface temperature decay subject to an surface impulse heating for Al on Si with and without the effects of electrons]{The surface temperature decay subject to an surface impulse heating for Al on Si with (solid blue line) and without (dashed red line) the effects of electrons. After 100 ps, the heat is dominated by phonons and there is little contribution from the electrons. Therefore, the electrons have negligible effects on the signal on the timescale relevant to the heat conduction across interfaces.}
\label{fig:ElectronJustification}
\end{figure*}

We compare the surface temperature responses to a heat impulse with and without the effects of electrons. As shown in Fig.~\ref{fig:ElectronJustification}, due to strong electron-phonon coupling, electrons only affect the heat conduction shortly after the absorption of a heat pulse. After the first 100 ps, the heat conduction is dominated by the phonons. Since a typical signal in a TDTR experiment is usually measured after 500 ps, whether heat is attributed to phonons or electrons in the metal has a negligible effect on the signal on the timescales that is interested in the experiments. Therefore, our neglect of electrons has no effect on our transmission coefficient measurement.

\section{Effects of mode conversion}\label{sec:ModeConversion}

When phonons cross an interface they can change their frequency, in an inelastic process, or polarization, known as mode conversion, which can influence thermal interface conductance.\cite{duda_assumption_2010} In our work, we do not consider inelastic scattering. We justify the neglect of inelastic scattering through the work of prior numerical studies, which have provided evidence that the phonon transmission between two slightly dissimilar crystalline solids is elastic.\cite{Murakami2014} Considering the phonon frequencies between Al and Si are very similar, there is no evidence that inelastic processes should play a role in the interfacial transport for Al/Si interfaces.

\begin{figure*}[t!]
\centering
\includegraphics[scale = 0.4]{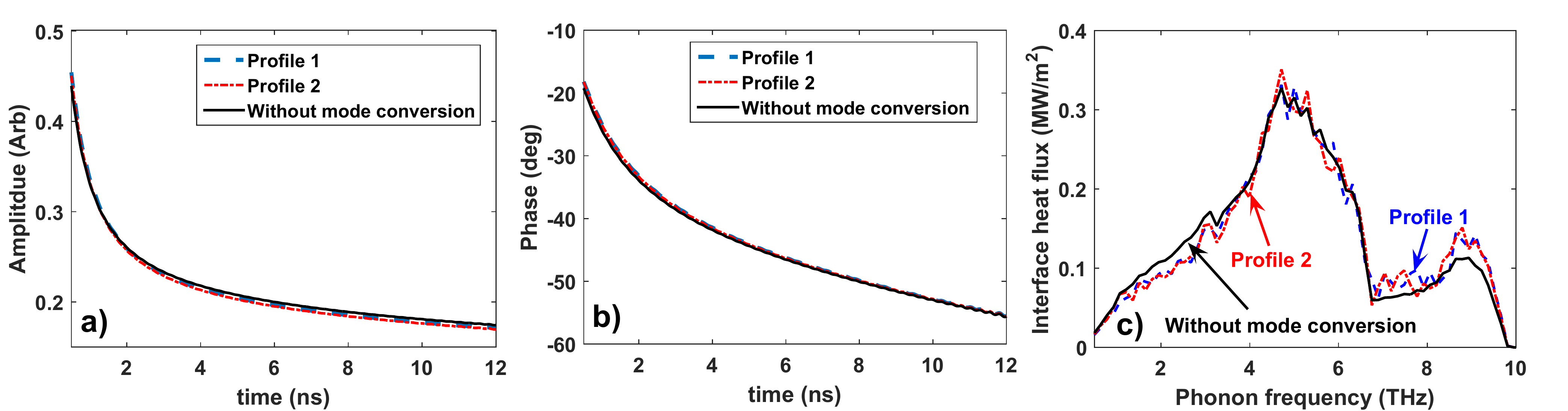}
\caption[TDTR surface temperature decay curves using different transmission coefficient profiles with two completely random partitions of transmitting modes to different polarizations and without mode conversion]{TDTR surface temperature decay curves along with (a) amplitude and (c) phase using different transmission coefficient profiles with two completely random partitions of transmitting modes to different polarizations (dashed and dash-dotted lines) and without mode conversion (solid line; assuming phonons maintain their polarization as they cross the interface). (c) Spectral interfacial heat flux versus phonon frequency predicted using different transmission coefficient profiles with mode conversion (dashed and dash-dotted lines) and without mode conversion. Mode conversion does not have an effect on the signal and conclusions beyond the uncertainties already considered in our model. }
\label{fig:ModeConversionJustification}
\end{figure*}

We have carefully examined the effect of mode conversion on our conclusions by rigorously including this process in our BTE model. To examine whether conversion between polarizations would affect the phonon transport across the interface, we conducted our BTE simulation assuming modes maintain their polarization after crossing the interface, or allowing them to change to any polarization while keeping the total transmission coefficient the same. Keeping the total transmission coefficient for a given polarization. Keeping $\sum_{j}T^{ij}_{12}(\omega)$ invariant, we randomly partitioned $\sum_{j}T^{ij}_{12}(\omega)$, $\sum_{j}R^{ij}_{12}(\omega)$ and $\sum_{j}R^{ij}_{21}(\omega)$ into two different combinations of $T^{ij}_{12}(\omega)$, $R^{ij}_{12}(\omega)$ and $R^{ij}_{21}(\omega)$;in other words, randomly between all the polarizations on the opposite side of the interface. In Fig.~\ref{fig:ModeConversionJustification}, we show that the surface temperature decay with and without conversion between polarization are essentially identical. Moreover, the spectral interfacial heat flux is also identical with and without conversion. Therefore, we conclude that mode conversion does not have an observable effect on the signal.

The reason that our measurement is not sensitive to mode conversion is that the polarizations in Si do not have extremely dissimilar mean free paths. As our measurement approach relies on the lack of scattering of some modes near the interface, the only way the mode conversion could affect our measurements would be if one polarization consistently changed to another polarization after transmitting through the interface with a drastically different mean free path than the original polarization. Our calculations clearly show that the difference in mean free paths between the polarizations is not sufficient to affect our calculations and hence have any effect on our conclusions. 

\section{Transmission coefficients for all polarizations}\label{sec:TransmissionCoeff}

\begin{figure*}[!h]
\centering
\includegraphics[scale = 0.3]{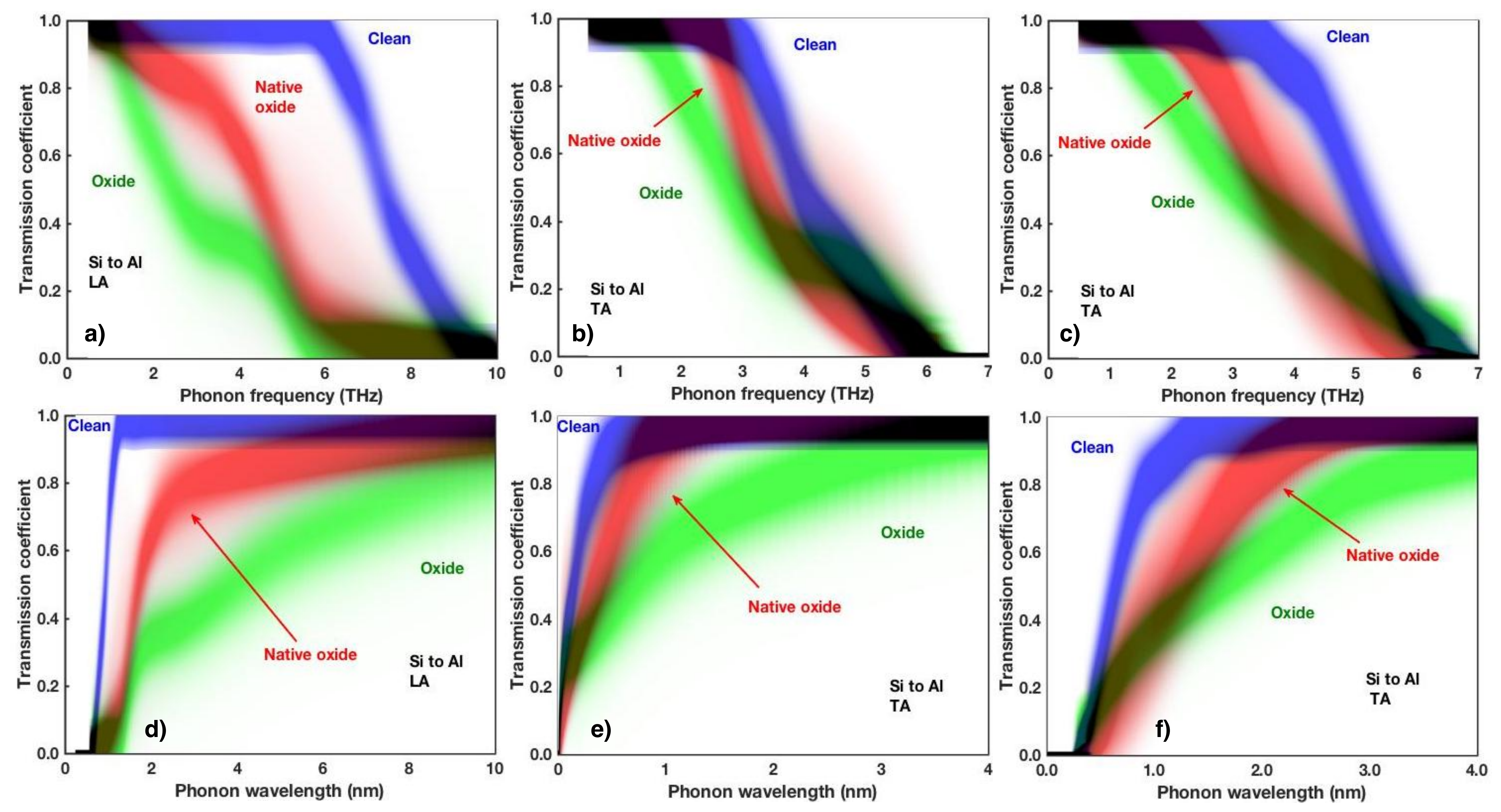}
\caption{Transmission coefficients from Si to Al versus (a)-(c) phonon frequency and (d)-(f) phonon wavelength for different polarizations measured from Al/Si sample with three different interfaces studied in this work. The intensity of the shaded region corresponds to the likelihood that the transmission coefficient possesses a given value. We emphasize that the transmission coefficients for the three polarizations are the only fitting parameters in our calculations.}
\label{fig:Trans_SitoAl}
\end{figure*}

\begin{figure*}[!h]
\centering
\includegraphics[scale = 0.36]{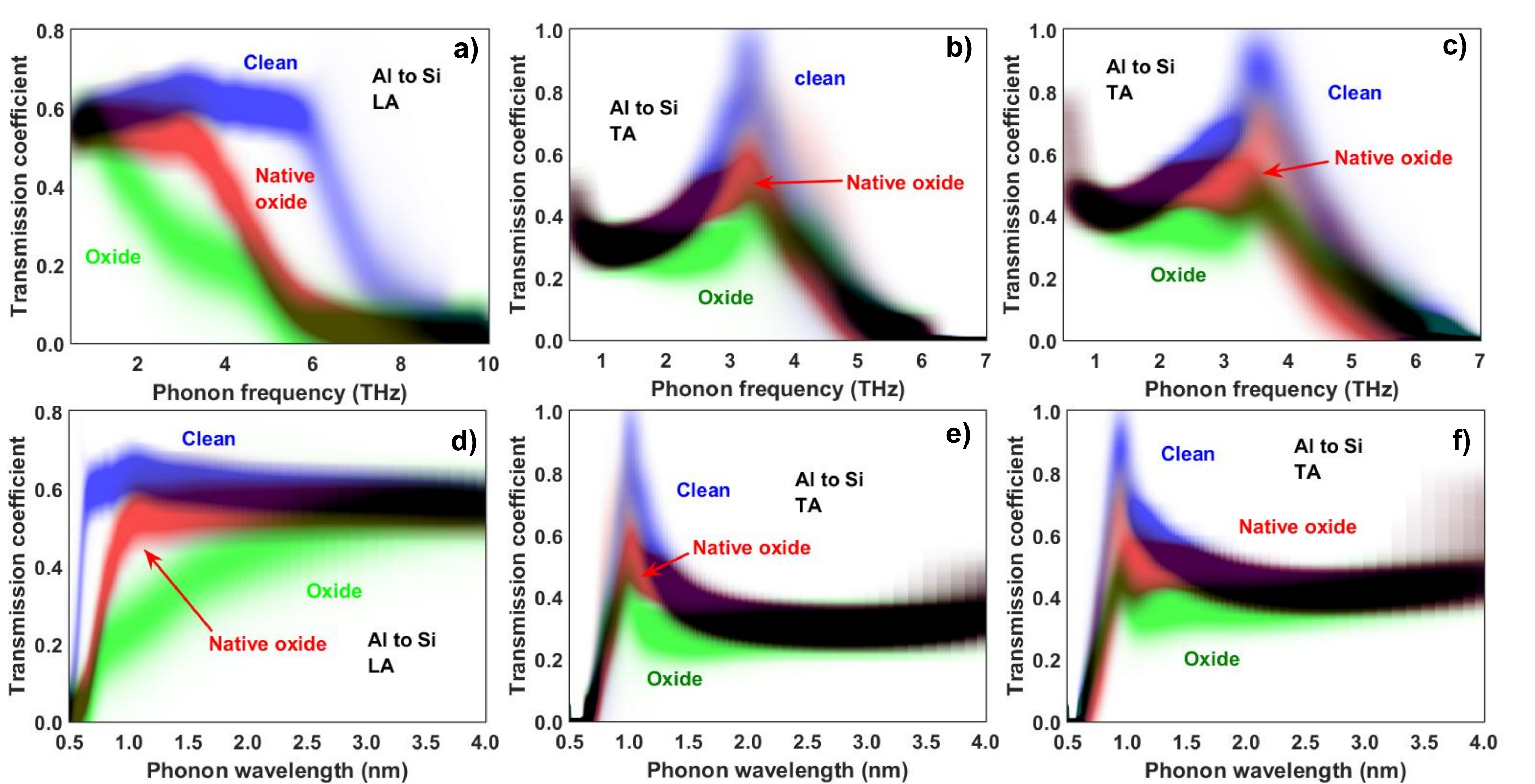}
\caption[Transmission coefficients from Al to Si for different polarizations measured from Al/Si sample with three different interfaces studied in this work]{Transmission coefficients from Al to Si versus (a)-(c) phonon frequency and (d)-(f) phonon wavelength for different polarizations measured from Al/Si sample with three different interfaces studied in this work, calculated using the measured values of the transmission coefficients from Si to Al shown in Fig.~\ref{fig:Trans_SitoAl}. The increase in transmission coefficient from Al to Si at phonon frequencies less than approximately 4 THz are due to the requirements of detailed balance. Specifically, these coefficients must follow the shape of the density of states since the coefficients from Si to Al are a constant value. These coefficients are determined by the principle of detailed balance and are not free parameters.}
\label{fig:Trans_AltoSi}
\end{figure*}

In the main text, we only show the transmission coefficient from Si to Al for longitudinal phonons for the three samples. Here, in Figs.~\ref{fig:Trans_SitoAl} and \ref{fig:Trans_AltoSi}, we plot the transmission coefficient profiles as a function of phonon frequency and wavelength from both sides of the materials for each polarization with a clean interface, with a native oxide layer and with a thermally grown oxide layer. The color intensity indicates the likelihood that a single transmission coefficient curve passing through a particular point at a given phonon frequency is able to simultaneously explain all of the experimental data. We emphasize that the only fitting parameters are the transmission coefficients from Si to Al for the three polarizations. All other transmission and reflection coefficients are determined from detailed balance and energy conservation.\cite{Minnich2011b}  

For the clean interface, the only constraint used in the fitting process is the smoothness of the profile. In particular, note that we do not enforce any type of monotonicity or shape requirement on the coefficients other than smoothness. For the native oxide interface, we additionally require that the transmission coefficients of the native oxide interface do not exceed the values for the clean interface. Similarly, the transmission coefficients of the thicker oxide interface should always be smaller than those of the native oxide interface.   

\clearpage

\section{TDTR data}\label{sec:TDTRdata}

\begin{figure*}[h!]
\centering
\includegraphics[scale = 0.35]{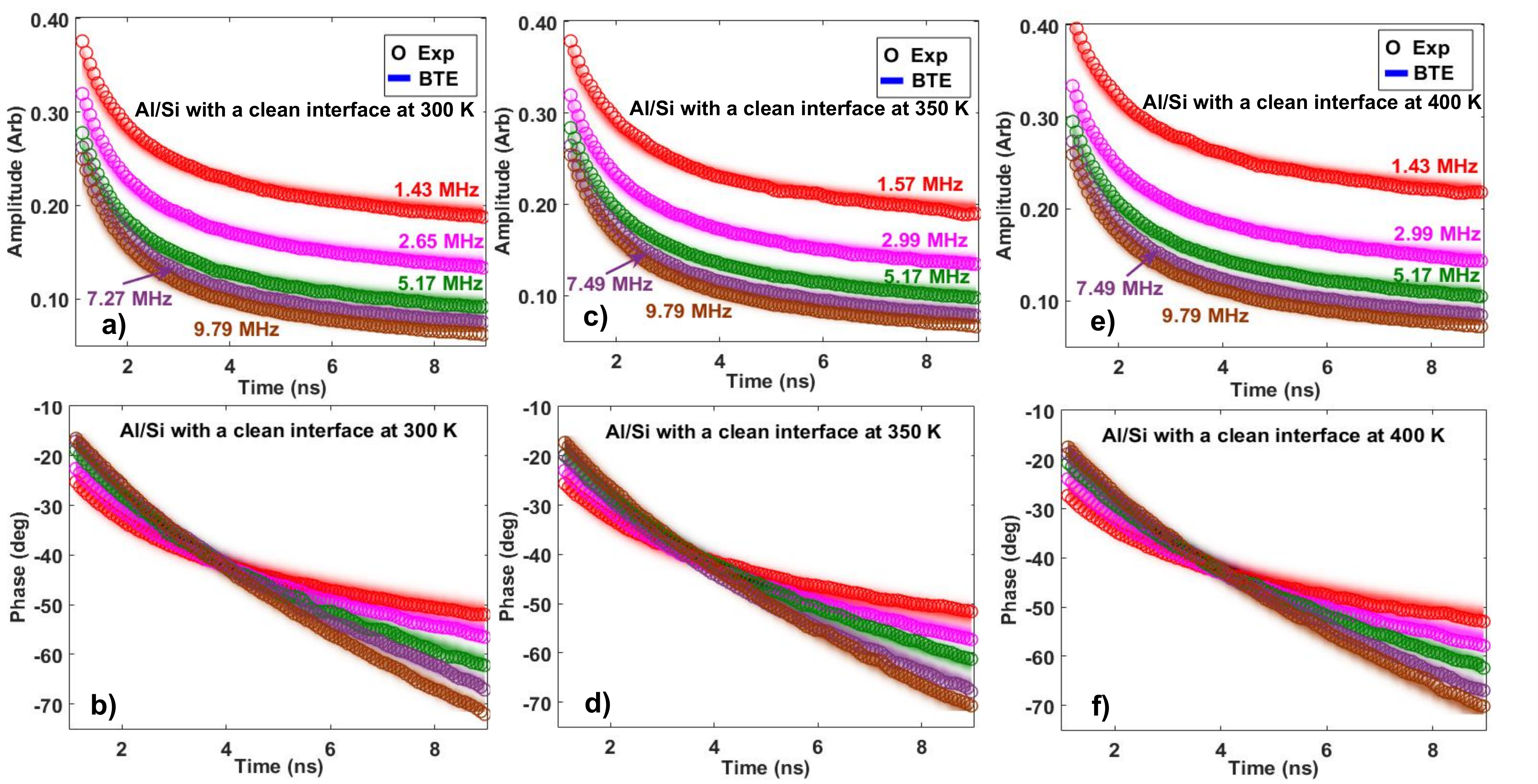}
\caption[Experimental TDTR data of an Al/Si sample with a clean interface at T = 300 K, 350 K, and 400 K at different modulation frequencies fit to the data from the BTE simulations]{Experimental TDTR data (symbols) of an Al/Si sample with a clean interface at T = 300 K, 350 K and 400 K at different modulation frequencies fit to the data from the BTE simulations (shaded regions), demonstrating excellent agreement between simulation and experiment at different temperatures.}
\label{fig:DecayCurves_VariousTemperatures}
\end{figure*}

\begin{figure*}
\centering
\includegraphics[scale = 0.35]{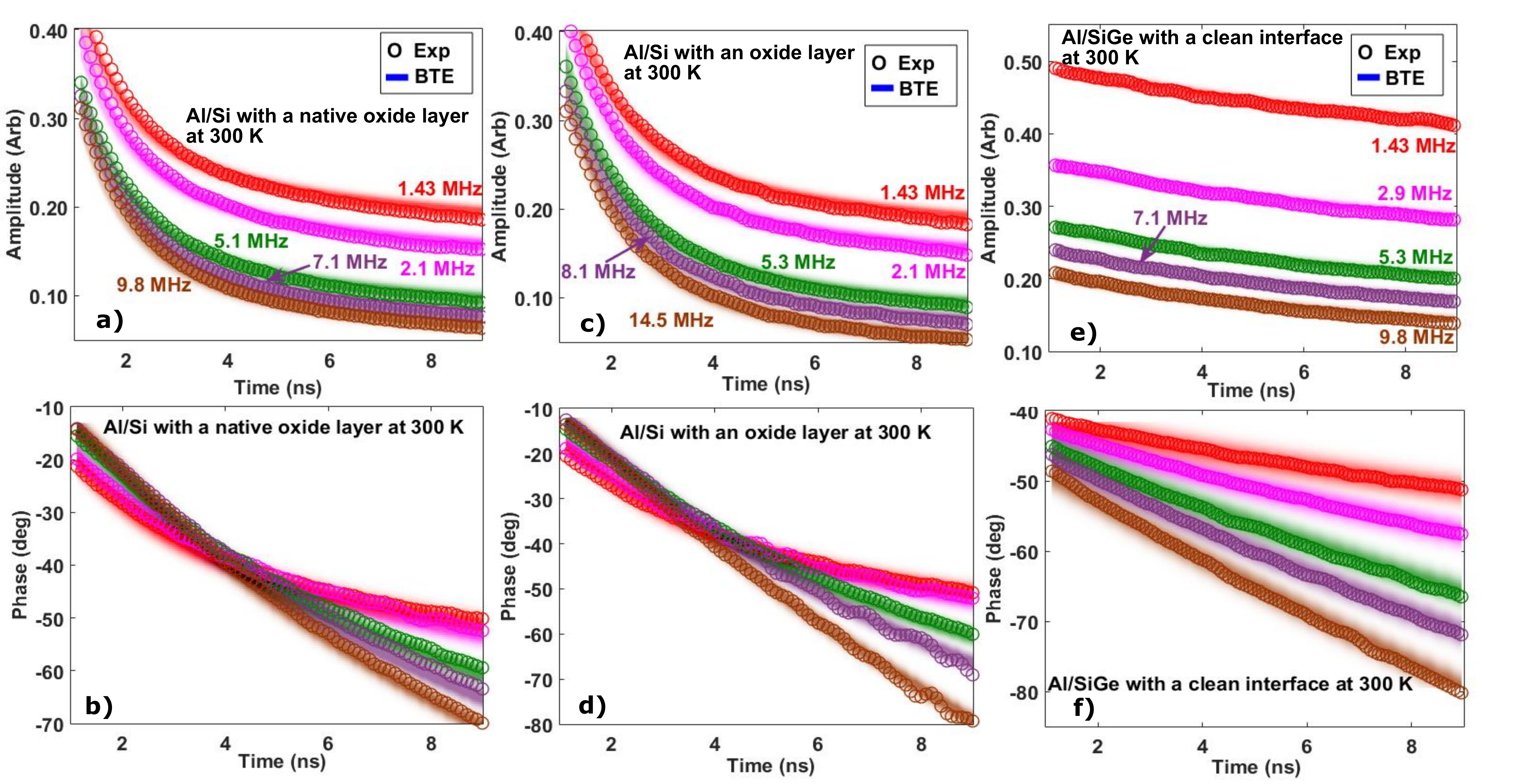}
\caption[Experimental TDTR data of an Al/Si sample with a native oxidized interface, an Al/Si sample with a thermally oxidized interface, and an Al/SiGe with a clean interface at T = 300 K at different modulation frequencies fit to the data from the BTE simulations]{Experimental TDTR data (symbols) of an Al/Si sample with a native oxidized interface, an Al/Si sample with a thermally oxidized interface, and an Al/SiGe with a clean interface at T = 300 K at different modulation frequencies fit to the data from the BTE simulations (shaded regions).}
\label{fig:DecayCurves_VariousInterfaces}
\end{figure*}

In Figs.~\ref{fig:DecayCurves_VariousTemperatures} \& \ref{fig:DecayCurves_VariousInterfaces}, we plot all the original raw data from the TDTR experiments used in the manuscript along with the BTE fitting results. In all the cases, we show excellent agreement between simulation and experiments. 

\clearpage

\section{Experimental details}\label{sec:ExperimentalDetails}

\subsection{Sample preparation}

Commercial high-purity natural Si (100) wafer and Si-Ge (1.5-2 at \% Ge) wafer (100) from MTI Corp. were used in the experiments. Before coating Al on the samples, three different surface conditions of the samples were prepared. First, the native oxide was removed with buffered HF acid to obtain a clean surface of Si and SiGe. After etching, the samples were immediately put into a vacuum chamber for Al deposition. Second, the native SiO$_2$ layer was left in place. No further treatment was taken for this condition before Al deposition. Finally, a thermally grown SiO$_2$ layer as fabricated by putting the Si samples into a tube furnace for three hours. The thickness of the native SiO$_2$ layer and thermally grown SiO$_2$ layer was measured by ellipsometry and TEM to be $\sim 1$ nm and $\sim 3.5$ nm, respectively. A thin film of Al was deposited on all samples using electron beam evaporator. The thickness of the Al transducer layer was 70 nm, measured by atomic force microscopy.

\subsection{TDTR measurements} 

The measurements are taken on two-tint TDTR. The details are available in Ref.~64.\nocite{Kang2008} The probe diameter is 10 $\mu$m and the pump diameter is 60 $\mu$m. Both beam sizes are measured using a home-built two-axis knife-edge beam profiler. With 60 $\mu$m pump heating size, the heat transfer problem can be treated as one-dimensional. All the measurements at $T = 300$ K are performed under ambient conditions, and the additional measurements at $T = 350$ and $400$ K are performed in an optical cryostat (JANIS ST-500) under high vacuum of $10^{-6}$ torr. 

\subsection{TEM images} 

The TEM samples were prepared by standard FIB lift-out technique in the dual beam FE-SEM/FIB (FEI Nova 600). To protect the top surface, a Pt layer with thickness $\sim$ 300 nm was deposited with electron beam evaporation followed by another Pt layer with thickness $\sim$ 3-4 $\mu$m by Ga ion beam. The lamella was cut parallel to the chip edge which was aligned to the wafer flat edge during initial cutting in TDTR sample preparation. As a result, the cutting surface normal was along (110) direction and all the TEM images were taken parallel to the Si (110) crystallographic zone axis. High resolution transmission electron microscopy (HRTEM) analyses were carried out in a FEI Tecnai TF-20 TEM/STEM at 200 kV. To avoid damage from the high energy electron beam, the beam exposure on region of interest was minimized especially at high magnification during operation.

\section{Ab-initio properties and modeling details}\label{sec:ModelingDetails}

\subsection{Point defect scattering in SiGe}\label{appsec:SiGe}
For SiGe, the mass difference scattering rate is calculated using the Tamura formula,\cite{Tamura1983} given by
\begin{equation}\label{eq:Tamura}
\tau^{-1} = \frac{\pi}{6}V_0m_0\omega^2D(\omega),
\end{equation}
where $\omega$ is phonon frequency, $D(\omega)$ is the phonon density of states per unit volume, and $V_0$ is the volume per atom. $m_0 = \sum_i f_i(1-m_i/\bar{m})^2$ is a measure of the mass disorder, $f_i$ and $m_i$ are the concentration and the atomic mass of species $i$, respectively, and $\bar{m}$ is the average mass for the given composition. The Tamura formula has been proven to effectively calculate the impurity scattering in SiGe with different Ge concentration.\cite{garg_role_2011} The values of all the constants in Eq.~\ref{eq:Tamura} are tabulated in Table~\ref{tab:constants}

We have sent the SiGe wafer to the third party, Thermotest, for bulk thermal conductivity measurements. The measured value, using transient plane source method on a bulk sample, is $50.7 \pm 0.5$ W/m-K. Using the measured value, we are able to obtain the Ge concentration to be about $\sim 2$ at \% based on calculations with the Tamura formula while the measured Ge concentration using Energy Dispersive X-ray Spectrometry is $\sim 1.5$ at \%, which gives SiGe thermal conductivity around $\sim 58$ W/m-K. These differences in atomic concentration have only a minimal effect on the transport calculations and have been incorporated in the uncertainty of BTE simulations in Figs.~\ref{fig:TDTR_clean_400K} (c) \& (d) of the main text. 

\subsection{Al thermal conductivity}\label{appsec:Alk}

\begin{figure*}[!t]
\centering
\includegraphics[scale = 0.55]{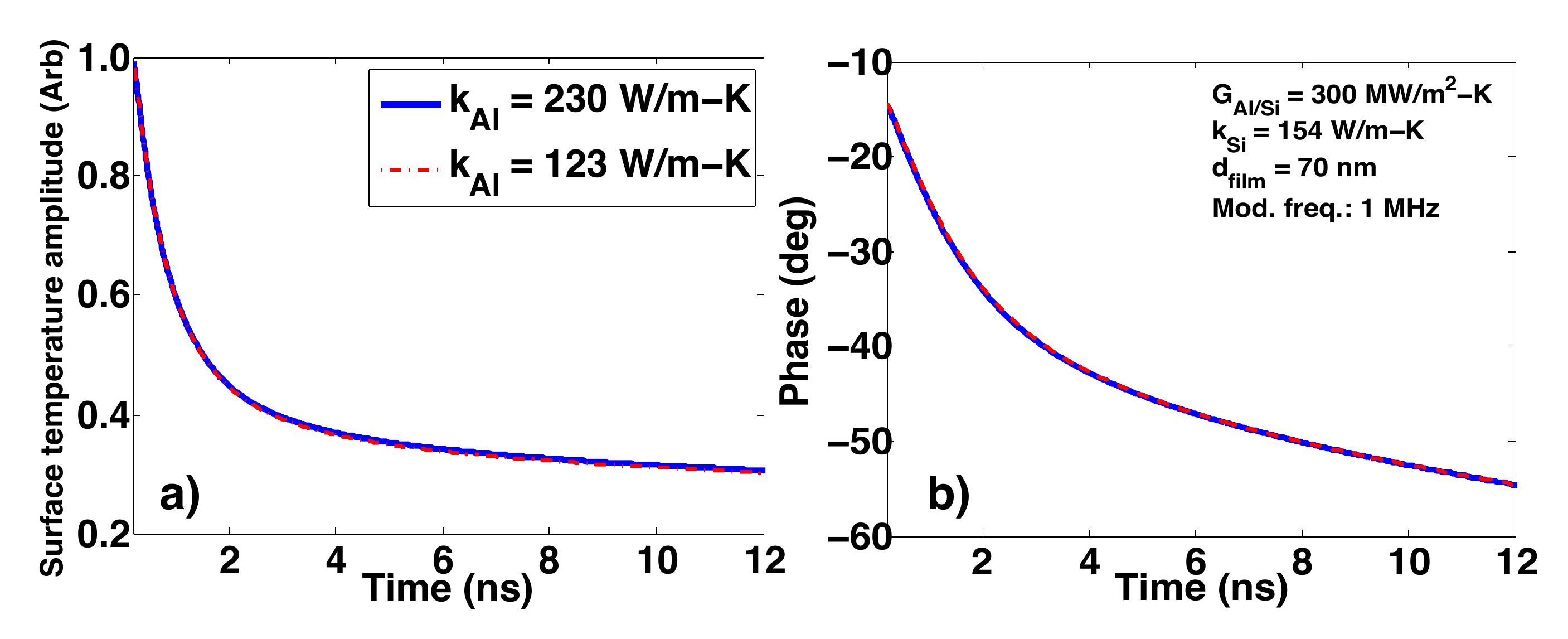}
\caption[Calculated transient surface temperature for Al on Si using a two-layer diffusive model with Al thermal conductivity to be 230 W/m-K and 123 W/m-K]{Calculated transient surface temperature (a) amplitude and (b) phase for Al on Si using a two-layer diffusive model with Al thermal conductivity to be 230 W/m-K (solid blue line) and 123 W/m-K (dash-dotted red line). The surface temperature response is not sensitive to the change of Al thermal conductivity from 230 W/m-K to 123 W/m-K.}
\label{fig:Alk_justification}
\end{figure*}

We assume a constant MFP for all modes in Al; the value $\Lambda_{Al} = 60$ nm is chosen to yield a lattice thermal conductivity $k \approx 123$ W/m-K so that no size effects in the thin film occur. Although the literature value of Al thermal conductivity is about $230$ W/m-K, we verified that the resulting surface temperature decay curves by using these two Al thermal conductivities in the TDTR diffusion model could not be distinguished as shown in Fig.~\ref{fig:Alk_justification}. Since the transmission coefficients are extracted by fitting our model to the data, if a parameter in the model has little effect on the results of the model, then it cannot affect the measured transmission coefficients. Here, we demonstrate that the calculations are completely insensitive to Al thermal conductivity, provided that it is larger than $\sim$ 30 W/m-K. Therefore, our choice of Al thermal conductivity has no impact on our results. 

The relaxation time for each mode is then obtained through $\tau_{\omega} = \Lambda_{Al}/v_{\omega}$. We also verified that the particular value of the Al MFP does not affect the results. Note that although the Al MPF is a constant, the dispersion of Al is directly from the first-principle calculations, and the transmission coefficients depend heavily on the density of states and phonon group velocity in both metal and substrate. Therefore, Al is still modeled with a spectral phonon BTE. 

\begin{table}
\caption[Constants appearing in the BTE models and the fitting process]{All the constants appearing in the BTE models and the fitting process are given in the following table.}\label{tab:constants}
\begin{center}
\begin{tabular}{|c|c|}
\hline
\multicolumn{2}{|c|}{\textbf{Bulk thermal properties}}\\
\hline
Al heat capacity (J/m$^3$-K): & $2.41 \times 10^6$ \\
\hline
Al lattice thermal conductivity (W/m-K): & 123\\
\hline
Al total thermal conductivity (W/m-K): & 230 \\
\hline
Si heat capacity (J/m$^3$-K): & $1.63 \times 10^6$\\
\hline
Si thermal conductivity (W/m-K): & 155\\
\hline
SiGe heat capacity (J/m$^3$-K): & $1.63 \times 10^6$\\
\hline
SiGe thermal conductivity (W/m-K): & 51\\
\hline
\multicolumn{2}{|c|}{\textbf{Electronic thermal properties in Al}}\\
\hline
Heat capacity (J/m$^3$-K): & $4.11 \times 10^4$\\
\hline
Thermal conductivity (W/m-K): & $203$\\
\hline
Electron-phonon coupling coefficient $g$ (W/m$^3$-K): &$2.1 \times 10^{17}$\\
\hline
\multicolumn{2}{|c|}{\textbf{Constants in Tamura formula}}\\
\hline
Volume per Si atom $V_0$ (nm$^3$): & 0.02 \\
\hline
Measure of the mass disorder $m_0$: & 0.0568\\
\hline
\multicolumn{2}{|c|}{\textbf{Transducer film thickness}}\\
\hline
Al/Si with a clean interface (nm): & 69\\
\hline
Al/SiGe with a clean interface (nm): & 72\\
\hline
Al/Si with a native oxidized interface (nm): & 70\\
\hline
Al/Si with a thermally-grown oxidized interface (nm): & 70\\
\hline
\multicolumn{2}{|c|}{\textbf{Other constants}}\\
\hline
Optical penetration depth $\delta$ (nm): & 10\\
\hline
Laser repetition frequency (MHz): & 76\\
\hline
\end{tabular}
\end{center}
\end{table}

\bibliographystyle{is-unsrt}
%\bibliography{Myref}

\begin{thebibliography}{10}
\ifx \showCODEN  \undefined \def \showCODEN #1{CODEN #1}  \fi
\ifx \showISBN   \undefined \def \showISBN  #1{ISBN #1}   \fi
\ifx \showISSN   \undefined \def \showISSN  #1{ISSN #1}   \fi
\ifx \showLCCN   \undefined \def \showLCCN  #1{LCCN #1}   \fi
\ifx \showPRICE  \undefined \def \showPRICE #1{#1}        \fi
\ifx \showURL    \undefined \def \showURL {URL }          \fi
\ifx \path       \undefined \input path.sty               \fi
\ifx \ifshowURL \undefined
     \newif \ifshowURL
     \showURLtrue
\fi

\bibitem{swartz_thermal_1989}
E.~T. Swartz and R.~O. Pohl.
\newblock Thermal boundary resistance.
\newblock {\em Rev. Mod. Phys.}, 61\penalty0 (3):\penalty0 605--668, July 1989.
\newblock \ifshowURL {\showURL
  \path|http://link.aps.org/doi/10.1103/RevModPhys.61.605|}\fi.

\bibitem{Cahill2014Review}
David~G. Cahill, Paul~V. Braun, Gang Chen, David~R. Clarke, Shanhui Fan,
  Kenneth~E. Goodson, Pawel Keblinski, William~P. King, Gerald~D. Mahan, Arun
  Majumdar, Humphrey~J. Maris, Simon~R. Phillpot, Eric Pop, and Li~Shi.
\newblock Nanoscale thermal transport. ii. 2003–2012.
\newblock {\em Applied Physics Reviews}, 1\penalty0 (1):\penalty0 011305, 2014.
\newblock \ifshowURL {\showURL
  \path|http://scitation.aip.org/content/aip/journal/apr2/1/1/10.1063/1.4832615|}\fi.

\bibitem{Swartz1989}
E.~T. Swartz and R.~O. Pohl.
\newblock Thermal boundary resistance.
\newblock {\em Rev. Mod. Phys.}, 61:\penalty0 605--668, Jul 1989.
\newblock \ifshowURL {\showURL
  \path|http://link.aps.org/doi/10.1103/RevModPhys.61.605|}\fi.

\bibitem{pettersson_theory_1990}
Sune Pettersson and G.~D. Mahan.
\newblock Theory of the thermal boundary resistance between dissimilar
  lattices.
\newblock {\em Phys. Rev. B}, 42\penalty0 (12):\penalty0 7386--7390, October
  1990.
\newblock \ifshowURL {\showURL
  \path|http://link.aps.org/doi/10.1103/PhysRevB.42.7386|}\fi.

\bibitem{Chen1998}
G.~Chen.
\newblock Thermal conductivity and ballistic-phonon transport in the
  cross-plane direction of superlattices.
\newblock {\em Phys. Rev. B}, 57:\penalty0 14958--14973, Jun 1998.
\newblock \ifshowURL {\showURL
  \path|http://link.aps.org/doi/10.1103/PhysRevB.57.14958|}\fi.

\bibitem{Ravichandran2014}
Jayakanth Ravichandran, Ajay~K. Yadav, Ramez Cheaito, Pim~B. Rossen, Arsen
  Soukiassian, S.~J. Suresha, John~C. Duda, Brian~M. Foley, Che-Hui Lee,
  Ye~Zhu, Arthur~W. Lichtenberger, Joel~E. Moore, David~A. Muller, Darrell~G.
  Schlom, Partick~E. Hopkins, Arun Majumdar, Ramamoorthy Ramesh, and Mark~A.
  Zurbuchen.
\newblock Crossover from incoherent to coherent phonon scattering in epitaxial
  oxide superlattices.
\newblock {\em Nature Materials}, 13:\penalty0 168--172, 2014.

\bibitem{Chen2013}
Peixuan Chen, N.~A. Katcho, J.~P. Feser, Wu~Li, M.~Glaser, O.~G. Schmidt,
  David~G. Cahill, N.~Mingo, and A.~Rastelli.
\newblock Role of surface-segregation-driven intermixing on the thermal
  transport through planar $\mathrm{Si}/\mathrm{Ge}$ superlattices.
\newblock {\em Phys. Rev. Lett.}, 111:\penalty0 115901, Sep 2013.
\newblock \ifshowURL {\showURL
  \path|http://link.aps.org/doi/10.1103/PhysRevLett.111.115901|}\fi.

\bibitem{SciencePaper}
Bed Poudel, Qing Hao, Yi~Ma, Yucheng Lan, Austin Minnich, Bo~Yu, Xiao Yan,
  Dezhi Wang, Andrew Muto, Daryoosh Vashaee, Xiaoyuan Chen, Junming Liu,
  Mildred~S. Dresselhaus, Gang Chen, and Zhifeng Ren.
\newblock High-thermoelectric performance of nanostructured bismuth antimony
  telluride bulk alloys.
\newblock {\em Science}, 320\penalty0 (5876):\penalty0 634--638, 2008.

\bibitem{Biswas2012}
K.~Biswas, J.~He, I.~V. Blum, C.~I. Wu, T.~P. Hogan, D.~N. Seidman, V.~P.
  Dravid, and M.~G Kanatzidis.
\newblock High-performance bulk thermoelectrics with all-scale hierarchical
  architectures.
\newblock {\em Nature}, 489:\penalty0 414--418, 2012.

\bibitem{Chiritescu2007}
C.~Chiritescu, D.~G. Cahill, N.~Nguyen, D.~Johnson, A.~Bodapati, P.~Keblinski,
  and P.~Zschack.
\newblock Ultralow thermal conductivity in disordered, layered wse2 crystals.
\newblock {\em Science}, 315:\penalty0 351--353, 2007.

\bibitem{Pop2010}
Eric Pop.
\newblock Energy dissipation and transport in nanoscale devices.
\newblock {\em Nano Research}, 3\penalty0 (3):\penalty0 147--169, 2010.
\newblock \showISSN{1998-0124}.
\newblock \ifshowURL {\showURL
  \path|http://dx.doi.org/10.1007/s12274-010-1019-z|}\fi.

\bibitem{Moore2014}
Arden~L. Moore and Li~Shi.
\newblock Emerging challenges and materials for thermal management of
  electronics.
\newblock {\em Materials Today}, 17\penalty0 (4):\penalty0 163 -- 174, 2014.
\newblock \showISSN{1369-7021}.
\newblock \ifshowURL {\showURL
  \path|http://www.sciencedirect.com/science/article/pii/S1369702114001138|}\fi.

\bibitem{Cho2015}
Jungwan Cho and Kenneth~E. Goodson.
\newblock Thermal transport: Cool electronics.
\newblock {\em Nature Materials}, 14:\penalty0 136--137, 2015.

\bibitem{Su2012}
Zonghui Su, Li~Huang, Fang Liu, Justin~P. Freedman, Lisa~M. Porter, Robert~F.
  Davis, and Jonathan~A. Malen.
\newblock Layer-by-layer thermal conductivities of the group iii nitride films
  in blue/green light emitting diodes.
\newblock {\em Applied Physics Letters}, 100\penalty0 (20):\penalty0 201106,
  2012.
\newblock \ifshowURL {\showURL
  \path|http://scitation.aip.org/content/aip/journal/apl/100/20/10.1063/1.4718354|}\fi.

\bibitem{Han2013}
Nam Han, Tran~Viet Cuong, Min Han, Beo~Deul Ryu, S.~Chandramohan, Jong~Bae
  Park, Ji~Hye Kang, Young-Jae Park, Kang~Bok Ko, Hee~Yun Kim, Hyum~Kyu Kim,
  Jae~Hyoung Ryu, Y.~S. Katharria, Chei-Jong Choi, and Chang-Hee Hong.
\newblock Improved heat dissiapation in gallium nitride light-emitting diodes
  with embedded graphene oxide pattern.
\newblock {\em Nature Communications}, 4:\penalty0 1452, 2013.

\bibitem{Yan2011}
Zhong Yan, Guanxiong Liu, Javed~M. Khan, and Alexander~A. Balandin.
\newblock Graphene quilts for thermal management of high-power gan transistors.
\newblock {\em Nature Communications}, 3:\penalty0 827, 2011.

\bibitem{klitsner_phonon_1987}
Tom Klitsner and R.~O. Pohl.
\newblock Phonon scattering at silicon crystal surfaces.
\newblock {\em Physical Review B}, 36\penalty0 (12):\penalty0 6551--6565,
  October 1987.
\newblock \showISSN{0163-1829}.
\newblock \ifshowURL {\showURL
  \path|http://link.aps.org/doi/10.1103/PhysRevB.36.6551|}\fi.

\bibitem{Swartz1987}
E.~T. Swartz and R.~O. Pohl.
\newblock Thermal resistance at interfaces.
\newblock {\em Applied Physics Letters}, 51\penalty0 (26), 1987.

\bibitem{Khalatnikov1952}
I.~M. Khalatnikov.
\newblock Teploobmen mezhdu tverdym telom i gelium — ii.
\newblock {\em Sov. Phys. JETP}, 22:\penalty0 687--704, 1952.

\bibitem{Little1959}
W.~A. Little.
\newblock The transport of heat between dissimilar solids at low temperatures.
\newblock {\em Canadian Journal of Physics}, 37\penalty0 (3):\penalty0
  334--349, 1959.
\newblock \ifshowURL {\showURL \path|http://dx.doi.org/10.1139/p59-037|}\fi.

\bibitem{Lyeo2006}
Ho-Ki Lyeo and David~G. Cahill.
\newblock Thermal conductance of interfaces between highly dissimilar
  materials.
\newblock {\em Phys. Rev. B}, 73:\penalty0 144301, Apr 2006.
\newblock \ifshowURL {\showURL
  \path|http://link.aps.org/doi/10.1103/PhysRevB.73.144301|}\fi.

\bibitem{Norris2009}
P.~M. Norris and P.~E. Hopkins.
\newblock Examining interfacial diffuse phonon scattering through transient
  thermoreflectance measurements of thermal boundary conductance.
\newblock {\em Journal of Heat Transfer}, 131:\penalty0 043207, 2009.

\bibitem{Cheaito2015}
Ramez Cheaito, John~T. Gaskins, Matthew~E. Caplan, Brian~F. Donovan, Brian~M.
  Foley, Ashutosh Giri, John~C. Duda, Chester~J. Szwejkowski, Costel
  Constantin, Harlan~J. Brown-Shaklee, Jon~F. Ihlefeld, and Patrick~E. Hopkins.
\newblock Thermal boundary conductance accumulation and interfacial phonon
  transmission: Measurements and theory.
\newblock {\em Phys. Rev. B}, 91:\penalty0 035432, Jan 2015.
\newblock \ifshowURL {\showURL
  \path|http://link.aps.org/doi/10.1103/PhysRevB.91.035432|}\fi.

\bibitem{Schmidt2010}
Aaron~J. Schmidt, Kimberlee~C. Collins, Austin~J. Minnich, and Gang Chen.
\newblock Thermal conductance and phonon transmissivity of metal–graphite
  interfaces.
\newblock {\em Journal of Applied Physics}, 107\penalty0 (10):\penalty0 104907,
  2010.
\newblock \ifshowURL {\showURL
  \path|http://scitation.aip.org/content/aip/journal/jap/107/10/10.1063/1.3428464|}\fi.

\bibitem{duda_role_2010}
John~C. Duda, Thomas~E. Beechem, Justin~L. Smoyer, Pamela~M. Norris, and
  Patrick~E. Hopkins.
\newblock Role of dispersion on phononic thermal boundary conductance.
\newblock {\em Journal of Applied Physics}, 108\penalty0 (7):\penalty0 073515,
  2010.
\newblock \showISSN{00218979}.
\newblock \ifshowURL {\showURL
  \path|http://scitation.aip.org/content/aip/journal/jap/108/7/10.1063/1.3483943|}\fi.

\bibitem{OBrien2012}
Peter~J. O'Brien, Sergei Shenogin, Jianxiun Liu, Philippe~K. Chow, Danielle
  Laurencin, P.~Hubert Mutin, Masashi Yamaguchi, Pawel Keblinski, and Ganpati
  Ramanath.
\newblock Bonding-induced thermal conductance enhancement at inorganic
  heterointerfaces using nanomolecular monolayers.
\newblock {\em Nature Materials}, 2012.
\newblock \showISSN{1476-1122}.
\newblock \ifshowURL {\showURL
  \path|http://www.nature.com.clsproxy.library.caltech.edu/nmat/journal/vaop/ncurrent/full/nmat3465.html|}\fi.

\bibitem{Losego2012}
M.~D. Losego, M.~E. Grady, N.~R. Sottos, D.~G. Cahill, and P.~V. Braun.
\newblock Effects of chemical bonding on heat transport across interfaces.
\newblock {\em Nature Materials}, 11:\penalty0 502--506, 2012.

\bibitem{Wang2011}
Zhaojie Wang, Joseph~E. Alaniz, Wanyoung Jang, Javier~E. Garay, and Chris
  Dames.
\newblock Thermal conductivity of nanocrystalline silicon: Importance of grain
  size and frequency-dependent mean free paths.
\newblock {\em Nano Letters}, 11\penalty0 (6):\penalty0 2206--2213, 2011.
\newblock \ifshowURL {\showURL \path|http://dx.doi.org/10.1021/nl1045395|}\fi.
\newblock PMID: 21553856.

\bibitem{Maiti1997}
A.~Maiti, G.D. Mahan, and S.T. Pantelides.
\newblock Dynamical simulations of nonequilibrium processes — heat flow and
  the kapitza resistance across grain boundaries.
\newblock {\em Solid State Communications}, 102\penalty0 (7):\penalty0 517 --
  521, 1997.
\newblock \showISSN{0038-1098}.
\newblock \ifshowURL {\showURL
  \path|http://www.sciencedirect.com/science/article/pii/S0038109897000495|}\fi.

\bibitem{Stevens2007}
Robert~J. Stevens, Leonid~V. Zhigilei, and Pamela~M. Norris.
\newblock Effects of temperature and disorder on thermal boundary conductance
  at solid–solid interfaces: Nonequilibrium molecular dynamics simulations.
\newblock {\em International Journal of Heat and Mass Transfer}, 50\penalty0
  (19–20):\penalty0 3977 -- 3989, 2007.
\newblock \showISSN{0017-9310}.
\newblock \ifshowURL {\showURL
  \path|http://www.sciencedirect.com/science/article/pii/S0017931007001342|}\fi.

\bibitem{Landry2009}
E.~S. Landry and A.~J.~H. McGaughey.
\newblock Thermal boundary resistance predictions from molecular dynamics
  simulations and theoretical calculations.
\newblock {\em Phys. Rev. B}, 80:\penalty0 165304, Oct 2009.
\newblock \ifshowURL {\showURL
  \path|http://link.aps.org/doi/10.1103/PhysRevB.80.165304|}\fi.

\bibitem{IhChoi2012}
Woon Ih~Choi, Kwiseon Kim, and Sreekant Narumanchi.
\newblock Thermal conductance at atomically clean and disordered
  silicon/aluminum interfaces: A molecular dynamics simulation study.
\newblock {\em Journal of Applied Physics}, 112\penalty0 (5):\penalty0 054305,
  2012.
\newblock \ifshowURL {\showURL
  \path|http://scitation.aip.org/content/aip/journal/jap/112/5/10.1063/1.4748872|}\fi.

\bibitem{Jones2013}
R.~E. Jones, J.~C. Duda, X.~W. Zhou, C.~J. Kimmer, and P.~E. Hopkins.
\newblock Investigation of size and electronic effects on kapitza conductance
  with non-equilibrium molecular dynamics.
\newblock {\em Applied Physics Letters}, 102\penalty0 (18):\penalty0 183119,
  2013.
\newblock \ifshowURL {\showURL
  \path|http://scitation.aip.org/content/aip/journal/apl/102/18/10.1063/1.4804677|}\fi.

\bibitem{Yang2013}
N.~Yang, T.~Luo, K.~Esfarjani, A.~Henry, Z.~Tian, J.~Shiomi, Y.~Chalopin,
  B.~Li, and G.~Chen.
\newblock Thermal interface conductance between aluminum and silicon by
  molecular dynamics simulations.
\newblock {\em Journal of Computational and Theoretical Nanoscience}, in press,
  2013.

\bibitem{Merabia2014}
Samy Merabia and Konstantinos Termentzidis.
\newblock Thermal boundary conductance across rough interfaces probed by
  molecular dynamics.
\newblock {\em Phys. Rev. B}, 89:\penalty0 054309, Feb 2014.
\newblock \ifshowURL {\showURL
  \path|http://link.aps.org/doi/10.1103/PhysRevB.89.054309|}\fi.

\bibitem{Liang2014}
Zhi Liang and Pawel Keblinski.
\newblock Finite-size effects on molecular dynamics interfacial
  thermal-resistance predictions.
\newblock {\em Phys. Rev. B}, 90:\penalty0 075411, Aug 2014.
\newblock \ifshowURL {\showURL
  \path|http://link.aps.org/doi/10.1103/PhysRevB.90.075411|}\fi.

\bibitem{Schelling2002b}
P.~K. Schelling, S.~R. Phillpot, and P.~Keblinski.
\newblock Phonon wave-packet dynamics at semiconductor interfaces by
  molecular-dynamics simulation.
\newblock {\em Applied Physics Letters}, 80\penalty0 (14), 2002.

\bibitem{Zhang2006}
W.~Zhang, T.S. Fisher, and N.~Mingo.
\newblock Simulation of interfacial phonon transport in si-ge heterostructure
  using an atomistic green’s function method.
\newblock {\em Journal of heat transfer}, 129:\penalty0 483--491, 2006.

\bibitem{Li2012b}
Xiaobo Li and Ronggui Yang.
\newblock Effect of lattice mismatch on phonon transmission and interface
  thermal conductance across dissimilar material interfaces.
\newblock {\em Phys. Rev. B}, 86:\penalty0 054305, Aug 2012.
\newblock \ifshowURL {\showURL
  \path|http://link.aps.org/doi/10.1103/PhysRevB.86.054305|}\fi.

\bibitem{Tian2014}
Zhiting Tian, Keivan Esfarjani, and Gang Chen.
\newblock Green's function studies of phonon transport across si/ge
  superlattices.
\newblock {\em Phys. Rev. B}, 89:\penalty0 235307, Jun 2014.
\newblock \ifshowURL {\showURL
  \path|http://link.aps.org/doi/10.1103/PhysRevB.89.235307|}\fi.

\bibitem{Huang2010}
Zhen Huang, Timothy~S. Fisher, and Jayathi~Y. Murthy.
\newblock Simulation of phonon transmission through graphene and graphene
  nanoribbons with a green’s function method.
\newblock {\em Journal of Applied Physics}, 108\penalty0 (9):\penalty0 094319,
  2010.
\newblock \ifshowURL {\showURL
  \path|http://scitation.aip.org/content/aip/journal/jap/108/9/10.1063/1.3499347|}\fi.

\bibitem{Hopkins2009}
Patrick~E. Hopkins, Pamela~M. Norris, Mikiyas~S. Tsegaye, and Avik~W. Ghosh.
\newblock Extracting phonon thermal conductance across atomic junctions:
  Nonequilibrium green’s function approach compared to semiclassical methods.
\newblock {\em Journal of Applied Physics}, 106\penalty0 (6):\penalty0 063503,
  2009.
\newblock \ifshowURL {\showURL
  \path|http://scitation.aip.org/content/aip/journal/jap/106/6/10.1063/1.3212974|}\fi.

\bibitem{Wilson2014b}
R.~Wilson and D.~Cahill.
\newblock Anisotropic failure of fourier theory in time-domain
  thermoreflectance experiments.
\newblock {\em Nature Communications}, 5\penalty0 (5075), 2013.

\bibitem{Capinski1996}
William~S. Capinski and Humphrey~J. Maris.
\newblock Improved apparatus for picosecond pump and probe optical
  measurements.
\newblock {\em Review of Scientific Instruments}, 67\penalty0 (8), 1996.

\bibitem{Schmidt2008}
Aaron~J. Schmidt, Xiaoyuan Chen, and Gang Chen.
\newblock Pulse accumulation, radial heat conduction, and anisotropic thermal
  conductivity in pump-probe transient thermoreflectance.
\newblock {\em Review of Scientific Instruments}, 79\penalty0 (11):\penalty0
  --, 2008.
\newblock \ifshowURL {\showURL
  \path|http://scitation.aip.org/content/aip/journal/rsi/79/11/10.1063/1.3006335|}\fi.

\bibitem{Koh2007}
Yee~Kan Koh and David~G. Cahill.
\newblock Frequency dependence of the thermal conductivity of semiconductor
  alloys.
\newblock {\em Phys. Rev. B}, 76:\penalty0 075207, Aug 2007.
\newblock \ifshowURL {\showURL
  \path|http://link.aps.org/doi/10.1103/PhysRevB.76.075207|}\fi.

\bibitem{Siemens2010}
Mark.~E. Siemens, Qing Li, Ronggui Yang, Keith~A. Nelson, Erik~H. Anderson,
  Murnane~Margaret M., and Henry~C. Kapteyn.
\newblock Quasi-ballistic thermal transport from nanoscale interfaces observed
  using ultrafast coherent soft x-ray beams.
\newblock {\em Nature Materials}, 9:\penalty0 29--30, 2010.

\bibitem{Minnich2011a}
A.~J. Minnich, J.~A. Johnson, A.~J. Schmidt, K.~Esfarjani, M.~S. Dresselhaus,
  K.~A. Nelson, and G.~Chen.
\newblock Thermal conductivity spectroscopy technique to measure phonon mean
  free paths.
\newblock {\em Phys. Rev. Lett.}, 107:\penalty0 095901, Aug 2011.
\newblock \ifshowURL {\showURL
  \path|http://link.aps.org/doi/10.1103/PhysRevLett.107.095901|}\fi.

\bibitem{Regner2012}
K.~Regner, D.~Sellan, Z.~Su, C.~Amon, A.~McGaughey, and J.~Malen.
\newblock Broadband phonon mean free path contributions to thermal conductivity
  measured using frequency-domain thermoreflectance.
\newblock {\em Nature Communications}, 4\penalty0 (1640), 2012.

\bibitem{Johnson2013}
Jeremy~A. Johnson, A.~A. Maznev, John Cuffe, Jeffrey~K. Eliason, Austin~J.
  Minnich, Timothy Kehoe, Clivia M.~Sotomayor Torres, Gang Chen, and Keith~A.
  Nelson.
\newblock Direct measurement of room-temperature nondiffusive thermal transport
  over micron distances in a silicon membrane.
\newblock {\em Phys. Rev. Lett.}, 110:\penalty0 025901, Jan 2013.
\newblock \ifshowURL {\showURL
  \path|http://link.aps.org/doi/10.1103/PhysRevLett.110.025901|}\fi.

\bibitem{Vermeersch2015a}
Bjorn Vermeersch, Amr M.~S. Mohammed, Gilles Pernot, Yee~Rui Koh, and Ali
  Shakouri.
\newblock Superdiffusive heat conduction in semiconductor alloys. ii. truncated
  l\'evy formalism for experimental analysis.
\newblock {\em Phys. Rev. B}, 91:\penalty0 085203, Feb 2015.
\newblock \ifshowURL {\showURL
  \path|http://link.aps.org/doi/10.1103/PhysRevB.91.085203|}\fi.

\bibitem{Cuffe2014}
John Cuffe, Jeffery~K. Eliason, Alexei~A. Maznev, Kimberlee~C. Collins,
  Jeremy~A. Johnson, Andrey Shchepetov, Mika Prunnila, Jouni Ahopelto, Clivia
  M.~S. Torres, Gang Chen, and Keith~A. Nelson.
\newblock Reconstructing phonon mean free path contributions to thermal
  conductivity using nanoscale membranes.
\newblock {\em arXiv:1408.6747}, 2014.

\bibitem{Maznve2011}
A.~A. Maznev, Jeremy~A. Johnson, and Keith~A. Nelson.
\newblock Onset of nondiffusive phonon transport in transient thermal grating
  decay.
\newblock {\em Phys. Rev. B}, 84:\penalty0 195206, Nov 2011.
\newblock \ifshowURL {\showURL
  \path|http://link.aps.org/doi/10.1103/PhysRevB.84.195206|}\fi.

\bibitem{Minnich2011b}
A.~J. Minnich, G.~Chen, S.~Mansoor, and B.~S. Yilbas.
\newblock Quasiballistic heat transfer studied using the frequency-dependent
  boltzmann transport equation.
\newblock {\em Phys. Rev. B}, 84:\penalty0 235207, Dec 2011.
\newblock \ifshowURL {\showURL
  \path|http://link.aps.org/doi/10.1103/PhysRevB.84.235207|}\fi.

\bibitem{Wilson2013}
R.~B. Wilson, Joseph~P. Feser, Gregory~T. Hohensee, and David~G. Cahill.
\newblock Two-channel model for nonequilibrium thermal transport in pump-probe
  experiments.
\newblock {\em Phys. Rev. B}, 88:\penalty0 144305, Oct 2013.
\newblock \ifshowURL {\showURL
  \path|http://link.aps.org/doi/10.1103/PhysRevB.88.144305|}\fi.

\bibitem{Regner2014}
K.~T. Regner, A.~J.~H. McGaughey, and J.~A. Malen.
\newblock Analytical interpretation of nondiffusive phonon transport in
  thermoreflectance thermal conductivity measurements.
\newblock {\em Phys. Rev. B}, 90:\penalty0 064302, Aug 2014.
\newblock \ifshowURL {\showURL
  \path|http://link.aps.org/doi/10.1103/PhysRevB.90.064302|}\fi.

\bibitem{Koh2014}
Yee~Kan Koh, David~G. Cahill, and Bo~Sun.
\newblock Nonlocal theory for heat transport at high frequencies.
\newblock {\em Phys. Rev. B}, 90:\penalty0 205412, Nov 2014.
\newblock \ifshowURL {\showURL
  \path|http://link.aps.org/doi/10.1103/PhysRevB.90.205412|}\fi.

\bibitem{Maassen2015}
Jesse Maassen and Mark Lundstrom.
\newblock Steady-state heat transport: Ballistic-to-diffusive with fourier's
  law.
\newblock {\em Journal of Applied Physics}, 117\penalty0 (3):\penalty0 035104,
  2015.
\newblock \ifshowURL {\showURL
  \path|http://scitation.aip.org/content/aip/journal/jap/117/3/10.1063/1.4905590|}\fi.

\bibitem{Majumdar1993}
A.~Majumdar.
\newblock Microscale heat conduction in dielectric thin-films.
\newblock {\em J. heat transfer}, 115:\penalty0 7--16, 1993.

\bibitem{hua_semi-analytical_2015}
Chengyun Hua and Austin~J. Minnich.
\newblock Semi-analytical solution to the frequency-dependent {Boltzmann}
  transport equation for cross-plane heat conduction in thin films.
\newblock {\em Journal of Applied Physics}, 117\penalty0 (17):\penalty0 175306,
  May 2015.
\newblock \showISSN{0021-8979, 1089-7550}.
\newblock \ifshowURL {\showURL
  \path|http://scitation.aip.org/content/aip/journal/jap/117/17/10.1063/1.4919432|}\fi.

\bibitem{Hua2014b}
Chengyun Hua and Austin~J. Minnich.
\newblock Analytical green's function of the multidimensional
  frequency-dependent phonon boltzmann equation.
\newblock {\em Phys. Rev. B}, 90:\penalty0 214306, Dec 2014.
\newblock \ifshowURL {\showURL
  \path|http://link.aps.org/doi/10.1103/PhysRevB.90.214306|}\fi.

\bibitem{Gangbook}
Gang Chen.
\newblock {\em Nanoscale Energy Transport and Conversion}.
\newblock Oxford University Press, New York, 2005.

\bibitem{hua2015}
Chengyun Hua and Austin~J. Minnich.
\newblock Semi-analytical solution to the frequency-dependent boltzmann
  transport equation for cross-plane heat conduction in thin films.
\newblock {\em Journal of Applied Physics}, 117\penalty0 (17):\penalty0 --,
  2015.
\newblock \ifshowURL {\showURL
  \path|http://scitation.aip.org/content/aip/journal/jap/117/17/10.1063/1.4919432|}\fi.

\bibitem{lee_phonon_2013}
Jaeho Lee, Elah Bozorg-Grayeli, SangBum Kim, Mehdi Asheghi, H.-S.~Philip Wong,
  and Kenneth~E. Goodson.
\newblock Phonon and electron transport through {Ge}2sb2te5 films and
  interfaces bounded by metals.
\newblock {\em Applied Physics Letters}, 102\penalty0 (19):\penalty0 191911,
  May 2013.
\newblock \showISSN{0003-6951, 1077-3118}.
\newblock \ifshowURL {\showURL
  \path|http://scitation.aip.org/content/aip/journal/apl/102/19/10.1063/1.4807141|}\fi.

\bibitem{giri_mechanisms_2015}
Ashutosh Giri, John~T. Gaskins, Brian~F. Donovan, Chester Szwejkowski,
  Ronald~J. Warzoha, Mark~A. Rodriguez, Jon Ihlefeld, and Patrick~E. Hopkins.
\newblock Mechanisms of nonequilibrium electron-phonon coupling and thermal
  conductance at interfaces.
\newblock {\em Journal of Applied Physics}, 117\penalty0 (10):\penalty0 105105,
  March 2015.
\newblock \showISSN{0021-8979, 1089-7550}.
\newblock \ifshowURL {\showURL
  \path|http://scitation.aip.org/content/aip/journal/jap/117/10/10.1063/1.4914867|}\fi.

\bibitem{Cahill2004}
David~G. Cahill.
\newblock Analysis of heat flow in layered structures for time-domain
  thermoreflectance.
\newblock {\em Review of Scientific Instruments}, 75\penalty0 (12):\penalty0
  5119--5122, 2004.
\newblock \ifshowURL {\showURL
  \path|http://scitation.aip.org/content/aip/journal/rsi/75/12/10.1063/1.1819431|}\fi.

\bibitem{Li:2012g}
Xiaobo Li and Ronggui Yang.
\newblock {Effect of lattice mismatch on phonon transmission and interface
  thermal conductance across dissimilar material interfaces}.
\newblock {\em Physical Review B}, 86\penalty0 (5):\penalty0 054305, August
  2012.
\newblock \ifshowURL {\showURL
  \path|http://link.aps.org/doi/10.1103/PhysRevB.86.054305|}\fi.

\bibitem{Li:2012j}
Xiaobo Li and Ronggui Yang.
\newblock {Size-dependent phonon transmission across dissimilar material
  interfaces}.
\newblock {\em Journal of Physics: Condensed Matter}, 24\penalty0
  (15):\penalty0 155302, April 2012.
\newblock \showISSN{0953-8984}.
\newblock \ifshowURL {\showURL
  \path|http://iopscience.iop.org/0953-8984/24/15/155302|}\fi.

\bibitem{Broido2007}
D.~A. Broido, M.~Malorny, G.~Birner, Natalio Mingo, and D.~A. Stewart.
\newblock Intrinsic lattice thermal conductivity of semiconductors from first
  principles.
\newblock {\em Applied Physics Letters}, 91\penalty0 (23):\penalty0 231922,
  2007.
\newblock \ifshowURL {\showURL
  \path|http://scitation.aip.org/content/aip/journal/apl/91/23/10.1063/1.2822891|}\fi.

\bibitem{Esfarjani2011}
Keivan Esfarjani, Gang Chen, and Harold~T. Stokes.
\newblock Heat transport in silicon from first-principles calculations.
\newblock {\em Phys. Rev. B}, 84:\penalty0 085204, Aug 2011.
\newblock \ifshowURL {\showURL
  \path|http://link.aps.org/doi/10.1103/PhysRevB.84.085204|}\fi.

\bibitem{zhao_phonon_2009}
Hong Zhao and Jonathan~B. Freund.
\newblock Phonon scattering at a rough interface between two fcc lattices.
\newblock {\em Journal of Applied Physics}, 105\penalty0 (1):\penalty0 , 2009.
\newblock \ifshowURL {\showURL
  \path|http://scitation.aip.org/content/aip/journal/jap/105/1/10.1063/1.3054383|}\fi.

\bibitem{Hohensee2015}
Gregory~T. Hohensee, Michael~R. Fellinger, Dallas~R. Trinkle, and David~G.
  Cahill.
\newblock Thermal transport across high-pressure semiconductor-metal transition
  in si and ${\mathrm{si}}_{0.991}{\mathrm{ge}}_{0.009}$.
\newblock {\em Phys. Rev. B}, 91:\penalty0 205104, May 2015.
\newblock \ifshowURL {\showURL
  \path|http://link.aps.org/doi/10.1103/PhysRevB.91.205104|}\fi.

\bibitem{shen_ballistic_2014}
Meng Shen and Pawel Keblinski.
\newblock Ballistic vs. diffusive heat transfer across nanoscopic films of
  layered crystals.
\newblock {\em Journal of Applied Physics}, 115\penalty0 (14):\penalty0 144310,
  April 2014.
\newblock \showISSN{0021-8979, 1089-7550}.
\newblock \ifshowURL {\showURL
  \path|http://scitation.aip.org/content/aip/journal/jap/115/14/10.1063/1.4870940|}\fi.

\bibitem{duda_assumption_2010}
John~C. Duda, Patrick~E. Hopkins, Justin~L. Smoyer, Matthew~L. Bauer,
  Timothy~S. English, Christopher~B. Saltonstall, and Pamela~M. Norris.
\newblock On the {Assumption} of {Detailed} {Balance} in {Prediction} of
  {Diffusive} {Transmission} {Probability} {During} {Interfacial} {Transport}.
\newblock {\em Nanoscale and Microscale Thermophysical Engineering},
  14\penalty0 (1):\penalty0 21--33, March 2010.
\newblock \showISSN{1556-7265, 1556-7273}.
\newblock \ifshowURL {\showURL
  \path|http://www.tandfonline.com/doi/abs/10.1080/15567260903530379|}\fi.

\bibitem{Murakami2014}
Takuru Murakami, Takuma Hori, Takuma Shiga, and Junichiro Shiomi.
\newblock Probing and tuning inelastic phonon conductance across
  finite-thickness interface.
\newblock {\em Applied Physics Express}, 7\penalty0 (12):\penalty0 121801,
  2014.
\newblock \ifshowURL {\showURL
  \path|http://stacks.iop.org/1882-0786/7/i=12/a=121801|}\fi.

\bibitem{Kang2008}
Kwangu Kang, Yee~Kan Koh, Catalin Chiritescu, Xuan Zheng, and David~G. Cahill.
\newblock Two-tint pump-probe measurements using a femtosecond laser oscillator
  and sharp-edged optical filters.
\newblock {\em Review of Scientific Instruments}, 79\penalty0 (11):\penalty0
  --, 2008.
\newblock \ifshowURL {\showURL
  \path|http://scitation.aip.org/content/aip/journal/rsi/79/11/10.1063/1.3020759|}\fi.

\bibitem{Tamura1983}
Shin-ichiro Tamura.
\newblock Isotope scattering of dispersive phonons in ge.
\newblock {\em Phys. Rev. B}, 27:\penalty0 858--866, Jan 1983.
\newblock \ifshowURL {\showURL
  \path|http://link.aps.org/doi/10.1103/PhysRevB.27.858|}\fi.

\bibitem{garg_role_2011}
Jivtesh Garg, Nicola Bonini, Boris Kozinsky, and Nicola Marzari.
\newblock Role of {Disorder} and {Anharmonicity} in the {Thermal}
  {Conductivity} of {Silicon}-{Germanium} {Alloys}: {A} {First}-{Principles}
  {Study}.
\newblock {\em Physical Review Letters}, 106\penalty0 (4), January 2011.
\newblock \showISSN{0031-9007, 1079-7114}.
\newblock \ifshowURL {\showURL
  \path|http://link.aps.org/doi/10.1103/PhysRevLett.106.045901|}\fi.

\end{thebibliography}

\end{document}